\documentclass[acmsmall, screen]{acmart}
\usepackage{booktabs} 
\usepackage{makecell}
\usepackage{lipsum}
\usepackage[utf8]{inputenc}
\usepackage{array}
\usepackage{wrapfig}
\usepackage{multirow}
\usepackage{multicol}
\usepackage{tabularx}
\usepackage{pifont}
\usepackage{lineno}
\usepackage{balance}
\usepackage{xcolor,pifont}

\usepackage{soul}

\usepackage[linesnumbered, ruled,vlined]{algorithm2e}
\RequirePackage{pdflscape}
\usepackage{flushend}
\usepackage{placeins}
\usepackage{booktabs}
\usepackage{indentfirst}
\usepackage{fancybox}
\usepackage[most]{tcolorbox}
\usepackage{fontawesome}
\usepackage{xurl} 
\usepackage{mathtools}
\usepackage{MnSymbol}
\usepackage{rotating}
\usepackage{adjustbox}
\usepackage{colortbl}
\usepackage{color}
\usepackage{enumitem}
\newtcolorbox{mybox}[2][]
{colback = white, colframe = black, fonttitle = \bfseries,
    colbacktitle = gray, enhanced,
    attach boxed title to top left={yshift=-3mm, xshift=3mm},
    title=#2, #1}

\usepackage{framed}
\definecolor{Gray}{gray}{0.9}
\definecolor{shadecolor}{gray}{0.95}
\usepackage{tikz}
\usetikzlibrary{trees,positioning,shapes,shadows,arrows}
\tikzset{
  basic/.style  = {draw, text width=2cm, drop shadow, font=\sffamily, rectangle},
  root/.style   = {basic, rounded corners=2pt, thin, align=center, fill=white},
  level-2/.style = {basic, rounded corners=6pt, thin,align=center, fill=white, text width=3cm},
  level-3/.style = {basic, thin, align=center, fill=white, text width=1.8cm}
}
\newcommand{\todo}[1]{}
\renewcommand{\todo}[1]{{\color{red} TODO: {#1}}}
\newcommand{\fakesection}[2][1em]{\vspace{#1}\noindent\textit{\textbf{#2}}}

\usepackage[normalem]{ulem}

\begin{document}
\title{Unveiling the Role of ChatGPT in Software Development: Insights from Developer-ChatGPT Interactions on GitHub}

\author{Ruiyin Li}
\orcid{0000-0001-8536-4935}
\email{ryli_cs@whu.edu.cn}
\affiliation{
  \institution{School of Computer Science, Wuhan University}
  \city{Wuhan}
  \country{China}
}

\author{Peng Liang}
\orcid{0000-0002-2056-5346}
\authornote{Corresponding author}
\email{liangp@whu.edu.cn}
\affiliation{
  \institution{School of Computer Science, Wuhan University}
  \city{Wuhan}
  \country{China}
}

\author{Yifei Wang}
\email{whiten@whu.edu.cn}
\orcid{0000-0003-0100-6896}
\affiliation{
  \institution{School of Computer Science, Wuhan University}
  \city{Wuhan}
  \country{China}
}

\author{Yangxiao Cai}
\orcid{0009-0007-7892-6611}
\email{yangxiaocai@whu.edu.cn}
\affiliation{
  \institution{School of Computer Science, Wuhan University}
  \city{Wuhan}
  \country{China}
}

\author{Weisong Sun}
\orcid{0000-0001-9236-8264}
\email{weisong.sun@ntu.edu.sg}
\affiliation{
  \institution{Nanyang Technological University}
  \city{Singapore}
  \country{Singapore}
}

\author{Zengyang Li}
\orcid{0000-0002-7258-993X}
\email{zengyangli@ccnu.edu.cn}
\affiliation{
  \institution{School of Computer Science, Central China Normal University}
  \city{Wuhan}
  \country{China}
}

\acmJournal{TOSEM}
\acmVolume{0}
\acmNumber{0}
\acmArticle{0}
\acmMonth{0}

\renewcommand{\shortauthors}{Li et al.}

\begin{abstract}
The advent of Large Language Models (LLMs) has introduced a new paradigm in Software Engineering (SE), with generative AI tools like ChatGPT gaining widespread adoption among developers. While ChatGPT's potential has been extensively discussed, empirical evidence about how developers actually use LLMs' assistance in real-world practices remains limited. To bridge this gap, we conducted a large-scale empirical analysis of ChatGPT usage on GitHub, and we presented DevChat, a curated dataset of 2,547 publicly shared ChatGPT conversation links collected from GitHub between May 2023 and June 2024. Through comprehensively analyzing DevChat, we explored the characteristics of developer–ChatGPT interaction patterns and identified five key categories of developers' purposes of using developer-ChatGPT conversations during software development. Additionally, we investigated the dominant development-related activities in which ChatGPT is used, and presented a mapping framework that links GitHub data sources, development-related activities, and SE tasks. The findings show that interactions are typically short and task-focused (most are 1-3 turns); developers share conversations mainly to delegate tasks, resolve problems, and acquire knowledge, revealing five purpose categories; ChatGPT is most frequently engaged for \textit{Software Implementation} and \textit{Maintenance \& Evolution}; we identified 39 fine-grained SE tasks supported by ChatGPT, with \textit{Code Generation \& Completion} as well as \textit{Code modification \& Optimization} being the most prominent. Our study offers a comprehensive mapping of ChatGPT's applications in real-world software development scenarios and provides a foundation for understanding LLMs’ practical roles in software development.
\end{abstract}

\maketitle

\section{Introduction}\label{sec:Introduction}
The emergence of Large Language Models (LLMs) has fundamentally reshaped the landscape of Software Engineering (SE) practices~\cite{Hou2024LLM4SE}. They introduce a novel paradigm for software development, offering capabilities that can potentially revolutionize the field. As one of the most typical and representative LLM-based tools, ChatGPT~\cite{OpenAI}, launched by OpenAI in November 2022, has sparked widespread interest within both academia and industry, leading to numerous studies exploring its applicability in software development, such as code generation and refinement \cite{Sobania2023aab, Guo2024epc, Li2023fft}, code summarization and comment generation~\cite{Weisong2025LLM4CodeSum, Su2024dgscs}, and program repair~\cite{Quanjun2024CRLLM}. Although prior research has demonstrated ChatGPT's capabilities and evaluated its performance on specific tasks, there remains limited empirical understanding of \textit{how} developers actually use ChatGPT in real-world software development workflows, and for \textit{what purposes} and \textit{tasks} it is leveraged in daily practice. 

A key challenge in studying real-world ChatGPT usage is the lack of observable traces that reflect developers' authentic practices. In May 2023, OpenAI released a new feature allowing users to share links within their ChatGPT interactions~\cite{ChatGPT_link}. This feature promotes the sharing of knowledge generated by ChatGPT and enhances team collaboration through shared developer-ChatGPT interactions. Developers create a unique record of ChatGPT-assisted development by embedding these links in GitHub workflows. As a result, this feature has further boosted ChatGPT's popularity among developers, surpassing other LLMs in developers' communities. As evidenced by our trial search, before selecting ChatGPT as our target LLM, we conducted a preliminary search on GitHub for other LLMs that support shared conversation links, such as Gemini~\cite{Gemini}
, and found significantly fewer shared links. Consequently, we decided to use ChatGPT as the focus of this study, which is the most popular and widely used LLM on GitHub.

Recent work has begun to explore such developer–ChatGPT interactions. For example, Xiao \textit{et al}. \cite{Xiao2024DevGPT} shared a dataset of developer-ChatGPT conversations used on GitHub, collected from July to October 2023, which ignited interest in studying ChatGPT's practical use. However, due to its relatively small size and temporal coverage, the dataset is not sufficient to support our comprehensive understanding of the prevalence and usability of ChatGPT in software development. Compared to the developer survey of ChatGPT usage conducted by Vaillant \textit{et al}. \cite{Vaillant2024ChatGPTsurvey}, our work provides empirical evidence of actual usage through observable ChatGPT links that developers proactively share. These links constitute concrete artifacts of ChatGPT usage, reflecting instances where developers considered the interactions sufficiently valuable to reference or reuse within development workflows. As such, our work offers complementary insights to perception- and recall-based survey studies by grounding the analysis in actually occurring usage data. We acknowledge that shared ChatGPT links represent only a subset of ChatGPT's overall usage in practice. Nevertheless, the links offer a valuable observation window into real-world developers' practices, capturing interactions that developers deem noteworthy or relevant enough to externalize. While this study does not aim to assess the direct impact of these shared conversations on development outcomes, it focuses on characterizing how ChatGPT is leveraged in developers' workflows through these observable usage traces. In this work, we mined and analyzed such shared and available ChatGPT links on GitHub to investigate the current landscape and patterns of ChatGPT usage in software development. For brevity, we henceforth refer to ``publicly shared ChatGPT usage'' simply as ``the usage of ChatGPT'' or ``ChatGPT usage'' in this paper. 

Additionally, compared to previous studies published in the Mining Challenge track of MSR 2024 \cite{MSR2024} that investigated the developer-ChatGPT interactions from certain perspectives (e.g., code generation, prompt interactions), our work comprehensively analyzed the ChatGPT usage during the software development process on GitHub through the shared and available ChatGPT links, including characterizing developer-ChatGPT \textit{interactions}, uncovering developers' \textit{motivations}, analyzing the development-related \textit{activities} developers use ChatGPT for, mapping GitHub data sources, development-related activities, and concrete SE \textit{tasks} performed with ChatGPT's assistance. The integration of ChatGPT into software development workflows is still in its early stages, but it has already shown great promise in improving developer productivity and code quality~\cite{Vaillant2024ChatGPTsurvey}. 



Therefore, our \textbf{goal} is to investigate how developers utilize ChatGPT in their daily routines to better anticipate its broader impact on the SE field. To this end, we conducted a large-scale empirical investigation into shared ChatGPT links, offering insights into LLM-assisted software development. Specifically, we presented \textbf{DevChat}, a curated dataset of 2,547 unique shared ChatGPT conversations collected from GitHub between May 2023 and June 2024. These conversations are drawn from five major GitHub data sources (i.e., Code, Issues, Commits, Pull Requests, and Discussions), capturing a diverse range of development scenarios. Leveraging DevChat, we conducted an in-depth analysis of how developers utilize ChatGPT during software development.

Our analysis reveals several \textbf{key findings}: Developer–ChatGPT interactions are typically involving short, low-turn conversations (most are 1-3 turns); Developers' purposes of using ChatGPT links on GitHub were identified and categorized into five categories, with \textit{Task Delegation} being the dominant purpose; We categorized the development-related activities within the shared ChatGPT conversations into two primary groups: \textit{Lifecycle Activities} and \textit{Supporting Activities}, with \textit{Software Implementation}, \textit{Software Maintenance and Evolution} emerged as the dominant activities; By systematically mapping data sources, development-related activities, and fine-grained SE tasks from shared ChatGPT links, our analysis reveals that ChatGPT's practical impact is predominantly concentrated in later SDLC phases, while its adoption in design tasks remains limited, with \textit{Code Generation \& Completion} and \textit{Code Modification \& Optimization} being the most prevalent. These findings offer valuable empirical evidence for guiding future research and the development of AI-assisted tools, contributing to a deeper understanding of the evolving AI-empowered development paradigm.

The main \textbf{contributions} of our study encompass:
\begin{itemize}
    \item \textbf{Dataset for Investigating Shared ChatGPT Links Usages}: We curated \textit{DevChat}, a dataset comprising 2,547 unique shared ChatGPT conversations on GitHub. Following deduplication and manual review, the cleaned dataset is available at [27]. This dataset enables future research into the practical applications of ChatGPT in software development.
    \item \textbf{Empirical Evidence of ChatGPT Usages on GitHub}: We presented the characteristics of ChatGPT usages on GitHub (RQ1); We identified and categorized developers' purposes of using developer-ChatGPT conversations during development in practice (RQ2); We examined the development-related activities for which developers shared ChatGPT links to support their development workflows (RQ3); We established a mapping relationship among GitHub data sources, development-related activities, and SE tasks associated with these shared ChatGPT links, providing a comprehensive view of ChatGPT's role in practical development scenarios (RQ4).
    \item \textbf{Insights for Practitioners and Researchers}: Based on our findings, we distilled concrete, evidence-based insights for practice and research, including context-aware integration of ChatGPT into development workflows, prioritizing tool support for dominant usage scenarios, and identifying underexplored SE areas where ChatGPT adoption remains limited. These insights provide evidence-based guidance for the responsible and effective use of ChatGPT.
\end{itemize}

The remainder of this paper is organized as follows: Section~\ref{sec:RelatedWork} introduces related studies of this work. Section~\ref{sec:Study Design} elaborates on the process of our study design and the research questions. Section~\ref{sec:Results} describes the study results and our findings, and the discussions and implications of this study are presented in Section~\ref{sec:Discussions & Implications}. Section~\ref{sec:Threats} examines the threats to validity. Section~\ref{sec:Conclusions} summarizes this study and outlines the future work.

\section{Related Work}\label{sec:RelatedWork}

\subsection{Large Language Models for Software Engineering}\label{sec:LLM4SE}
In recent years, groundbreaking advancements in Natural Language Processing (NLP) have fueled a meteoric rise in the performance and adoption of Large Language Models (LLMs). The breakthroughs allow LLMs to scale up the model sizes and significantly enhance their capacity when the parameter scale surpasses a certain threshold \cite{Hou2024LLM4SE}. As a result, LLMs have become increasingly powerful and versatile, driving innovation and efficiency across various fields and applications. These LLMs, such as OpenAI's GPT series models (e.g., GPT-1 $\sim$ GPT-4o), Google's Gemini~\cite{Gemini}, Meta's Llama~\cite{LLaMA}, and Anthropic's Claude~\cite{Claude}, are built on the Transformer architecture \cite{Vaswani2017Transformer} to understand and generate text through the training on extensive datasets comprising diverse linguistic patterns and contexts.

LLMs are extensively utilized throughout various phases of the Software Development Lifecycle (SDLC), including requirements engineering, software design, software development, quality assurance, maintenance, and management \cite{Hou2024LLM4SE}. The most predominant phase is software development, and code generation and program repair being the most prevalent tasks for LLMs \cite{Hou2024LLM4SE}. 
For example, as a code generation tool empowered by LLM, GitHub Copilot \cite{Copilot} received much attention from software developers after its release. Vaithilingam \textit{et al}. \cite{Vaithilingam2022Copilot} conducted a user study with 24 participants on the usability of GitHub Copilot. They found that most participants preferred using Copilot, as Copilot provided a useful starting point for programming tasks and saved them the effort of searching code snippets online. 
Sergeyuk \textit{et al}. \cite{Sergeyuk2025AICoding} presents an empirical study on the practical use of AI-based coding assistants, examining developers' adoption patterns, perceptions, and challenges. Their findings highlight usage practices and limitations of AI-based coding assistants, offering insights that complement research on the broader impacts of LLM-based tools. 
Liang \textit{et al}. \cite{Liang2024surveyAI} performed an exploratory study to investigate the usability of LLM-assisted programming tools, such as GitHub Copilot, with 410 programmers. Their results indicate that developers are motivated to use AI programming assistants (e.g., GitHub Copilot) because these tools help reduce keystrokes, speed up tasks, and recall syntax, but there is still a gap between developers' needs and the tools' output. 
Jin \textit{et al}. \cite{jin2023inferfix} introduced an end-to-end program repair framework, InferFix, to fix critical security and performance bugs in Java and C\#. By combining a static analyzer with a fine-tuned LLM, InferFix shows better performance than LLM baselines on patch generation in Java and C\#. 
Moreover, a recent study of Poldrack \textit{et al}. \cite{Poldrack2023aac} reported the experimental results of AI-assisted coding with GPT4, including performing required coding tasks, refactoring, and improving the quality of existing code. Their results show that although AI coding tools are powerful, human involvement is still necessary to ensure the validity and accuracy of the results. 


\subsection{ChatGPT in Software Development}\label{sec:UsingChatGPT}
Researchers prefer ChatGPT over other LLMs and LLM-based applications due to its computational efficiency, adaptability to various tasks, and potential cost-effectiveness~\cite{Hou2024LLM4SE}. As a result, the use of ChatGPT in software development encompasses a wide spectrum of applications. It ranges from generating code snippets and debugging to providing documentation, and ChatGPT proves to be a versatile tool for various stages of the development process. 
Hao \textit{et al}. \cite{Hao2024esd} conducted an empirical study to investigate the role of ChatGPT in collaborative coding. They specially focused on developers' shared conversations with ChatGPT in GitHub pull requests and issues, based on the DevGPT dataset \cite{Xiao2024DevGPT}. Their work uncovered how developers seek ChatGPT's assistance across 16 types of inquiries; the most frequently employed inquiries involve code generation, conceptual understanding, and others. 
Sobania \textit{et al}. \cite{Sobania2023aab} evaluated and analyzed the automatic bug-fixing performance of ChatGPT. Their findings show that ChatGPT's program repair performance is competitive with the results achieved with common deep learning approaches like CoCoNut~\cite{Lutellier2020coconut} and Codex~\cite{Prenner2022Codex}, and notably better than the results reported for the standard program repair approaches.

Besides, certain concerns about the correctness of ChatGPT-generated information have been raised. 
Kabir \textit{et al}. \cite{Kabir2024soo} empirically investigated the characteristics of ChatGPT answers to 517 programming questions in Stack Overflow alongside manual analysis, linguistic analysis, and user study. Their findings indicate that ChatGPT produces incorrect answers in more than 50\% of the cases, and manual validation identified a large number of conceptual and logical errors in the ChatGPT answers. 
Li \textit{et al}. \cite{Li2023fft} proposed a new paradigm for finding failure-inducing test cases through leveraging ChatGPT. They found that ChatGPT initially struggled with pinpointing the buggy code, especially when program versions had subtle differences. To address this weakness of ChatGPT, the authors designed a novel approach by leveraging ChatGPT's ability to infer expected behavior from erroneous programs and amplify the subtle code differences. Their experimental results show that the approach can significantly improve the accuracy of identifying fault-inducing test cases. 
Vaillant \textit{et al}. \cite{Vaillant2024ChatGPTsurvey} surveyed 207 developers to investigate the impact of ChatGPT on software quality, productivity, and job satisfaction. Their results show that around 70\% of developers perceive ChatGPT as improving productivity and job satisfaction.
Aguiar \textit{et al}. \cite{Aguiar2024ChatGPT} conducted an empirical study to analyze developers' use of ChatGPT in multi-language software development. They found that 70\% of the developer-ChatGPT conversations are seeking coding support regarding multiple programming languages, while 57\% of developers used ChatGPT as a tool to generate code in multiple languages.


\subsection{Conclusive Summary}\label{sec:ConclusiveSummary}

Table~\ref{T:Comparison} provides a structured overview of prior studies, the existing research gaps, and the distinctions between our work and previous efforts. The table is organized by four columns: \textit{Theme} (topics of related work), \textit{Previous Studies} (summarizations of previous studies), \textit{Gap} (what remains unexplored), and \textit{Comparison} (how our study fills these gaps).

While previous studies (see Table~\ref{T:Comparison}) have investigated the capabilities of LLMs and ChatGPT in software development, their focuses have largely been on task-specific evaluations, isolated interaction patterns, or model-centric performance metrics. Consequently, there is limited understanding of how developers leverage ChatGPT into everyday software development workflows and the broader range of activities supported by ChatGPT in practice.

Our study addresses these gaps by adopting a developer-centric perspective and analyzing real-world usage of ChatGPT. Our work provides a more comprehensive and nuanced understanding of ChatGPT usage across the software development lifecycle, revealing practical usage patterns, underlying developers' motivations, and specific types of SE tasks it supports.


\begin{table}[htb]
    \centering
    \footnotesize
    \caption{Gap and Comparison between our work and prior studies}\label{T:Comparison}
    \begin{tabular}[width=\textwidth]{|p{14mm}|p{40mm}|p{35mm}|p{34mm}|}\hline
        \textbf{Theme} & \textbf{Previous Studies} & \textbf{Gap} & \textbf{Comparison} \\\hline
        \textbf{The use of LLMs in SE} 
        & 
        LLMs have been used for various SE tasks, including automated code generation \cite{Jiang2024LLMCG, Sergeyuk2025AICoding}, code completion \cite{Husein2025LLMCC}, bug detection and repair \cite{Quanjun2024CRLLM}, and documentation synthesis \cite{Dvivedi2024LLMCDG}, demonstrating LLMs' potential to augment developer productivity and in software development. 
        & 
        Previous studies primarily adopt task-centric perspectives, focusing on model capabilities, performance comparisons, or qualitative assessments of output quality. While they provide valuable insights into what LLMs can do for specific SE tasks, comprehensive insights into how developers actually leverage LLMs into their daily development workflows remain unexplored. 
        & 
        Our work complements and extends existing LLM4SE research by shifting the focus from model-centric capabilities to developer-centric practices, providing a holistic empirical foundation for understanding the practical ChatGPT usage, SE task distribution assisted by ChatGPT, and potential areas for future AI-assisted development tool enhancements. \\\hline
        
        \textbf{The use of ChatGPT in software development}
        & 
        As one of the most representative LLMs, ChatGPT has been used in diverse SE tasks. For example, previous studies on the Mining Challenge Track of MSR 2024 \cite{MSR2024} primarily focused on four aspects: (1) \textit{Code Generation Quality and Developers' Perceptions}, such as analysis of the quality and correctness of ChatGPT-generated code (e.g., \cite{Siddiq2024QA}), comparison between ChatGPT-generated code and human-written alternatives (e.g., \cite{Rabbi2024}). (2) \textit{Developers' Behaviors and Practices}, such as developers' interactions with ChatGPT and practices in real-world scenarios (e.g., \cite{Mohamed2024cwa}). (3) \textit{Code Integration and Modification}: Integration and Modification of ChatGPT-generated code into projects (e.g., \cite{AlOmar2024ref}). (4) \textit{Prompt Interaction and Iterative Resolution}: Investigation of the iterative nature of prompting and its use in issue resolutions (e.g., \cite{Wu2024prompt}).
        & 
        While previous studies provide valuable insights into specific use of ChatGPT, they remain limited in scope, focusing on isolated tasks or interaction patterns rather than covering the comprehensive lifecycle of software development. In particular, prior work does not systematically analyze how developer-ChatGPT interactions reflect developers' practical usage across diverse development activities, their underlying motivations behind the developer-ChatGPT interactions, or the specific SE tasks facilitated by ChatGPT. 
        & 
        Our work comprehensively analyzed the sharing of ChatGPT links during the software development process on GitHub. Specifically, we investigated practical ChatGPT sharing in real-world applications based on our established large and cleaned dataset (i.e., DevChat \cite{onlinepackage_TOSEM}):
        \newlist{mylist}{itemize}{1}
        \setlist[mylist]{label=\textbullet, leftmargin=4pt, itemindent=4pt}
        \begin{mylist}
        \item Characterize \textit{how} developers interact with ChatGPT (RQ1).
        \item Uncover the \textit{motivations} behind the developer-ChatGPT interactions (RQ2).
        \item Analyze the development-related \textit{activities} for which developers share ChatGPT links (RQ3).
        \item Categorize the concrete software engineering \textit{tasks} performed by developers with the help of shared ChatGPT links (RQ4).
        \end{mylist}
        \\\hline
    \end{tabular}
\end{table}

\section{Study Design}\label{sec:Study Design}
\begin{figure}[hbtp]
	\centering
        \includegraphics[width=0.9\linewidth]{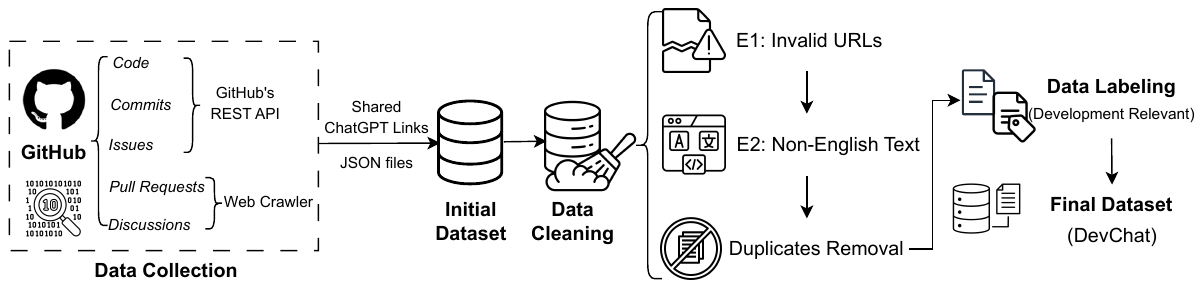}
	\caption{Overview of the DevChat dataset curation process}\label{F:Overview}
\end{figure}

In this section, we present the Research Questions (RQs) (Section~\ref{sec:RQs}) used to achieve the goal stated in the Introduction section (Section~\ref{sec:Introduction}) and provide an overview of the DevChat dataset curation process containing data collection (Section~\ref{sec:Data Collection}), data cleaning (Section~\ref{sec:Data Cleaning}), data labeling (Section~\ref{sec:Data Labeling}), and data analysis (Section~\ref{sec:Data Analysis}). Figure~\ref{F:Overview} shows an overview of the DevChat dataset curation process. The dataset goes through a multi-step process, which includes searching data from five sources on GitHub, filtering out invalid URLs, removing non-English text and duplicates. Then, the cleaned data is subsequently labeled before being compiled into the final curated dataset.

\subsection{Research Questions}\label{sec:RQs}

In this work, our \textbf{goal} aims to provide a detailed and comprehensive investigation into the usage of ChatGPT in software development. To this end, we formulated four Research Questions (RQs): RQ1 characterizes \textit{how} developers interact with ChatGPT, RQ2 uncovers the \textit{motivations} behind those interactions, RQ3 analyzes the development-related \textit{activities} for which developers share ChatGPT links, and RQ4 categorizes the concrete SE \textit{tasks} performed by developers with the help of shared ChatGPT links.

\begin{tcolorbox}[colback=gray!15, colframe=gray]
\textbf{RQ1: How do developers interact with ChatGPT during software development?}
\end{tcolorbox}
\textbf{Rationale}: RQ1 aims to analyze the distributional and usage characteristics of shared ChatGPT links on GitHub, such as the frequency and distribution of prompt turns within developer–ChatGPT interactions, and the contextual descriptions accompanying shared ChatGPT links. By analyzing these dimensions, RQ1 can uncover how developers leverage ChatGPT throughout the development process.


\begin{tcolorbox}[colback=gray!15, colframe=gray]
\textbf{RQ2: What are the purposes for which developers use ChatGPT based on the shared ChatGPT links?}
\end{tcolorbox}
\textbf{Rationale}: By examining the specific purposes for which developers use ChatGPT based on the shared ChatGPT links during software development, RQ2 seeks to provide nuanced insights into the developers' intentions in employing ChatGPT beyond the technical tasks involved. This purpose-oriented understanding complements task-level analysis and provides insight into how ChatGPT shapes collaboration and information exchange in software development.

\begin{tcolorbox}[colback=gray!15, colframe=gray]
\textbf{RQ3: Which development-related activities do developers share ChatGPT links for during software development?}
\end{tcolorbox}
\textbf{Rationale}: RQ3 is designed to map out a comprehensive landscape of ChatGPT usage during software development. By identifying and categorizing various development-related activities that are widely recognized in SWEBOK \cite{SWEBOKv04} and prior studies \cite{Hou2024LLM4SE, Fan2023LLMSurvey}, this RQ quantitatively examines the distribution of ChatGPT links across these activities, offering insights into ChatGPT's practical applications and impact on the software development process.

\begin{tcolorbox}[colback=gray!15, colframe=gray]
\textbf{RQ4: What tasks do developers handle with ChatGPT during software development?}
\end{tcolorbox}
\textbf{Rationale}: RQ4 delves into analyzing the fine-grained tasks that shared ChatGPT conversations engage in within broader development-related activities, such as \textit{debugging} within \textit{software maintenance and evolution}. By analyzing task-specific interactions, RQ4 seeks to systematically categorize the types of ChatGPT-assisted tasks within dedicated activities, and derive actionable insights for optimizing ChatGPT's utility and efficiency.



\subsection{Data Collection}\label{sec:Data Collection}
We curated our DevChat dataset~\cite{onlinepackage_TOSEM} by collecting conversation records of developers interacting with ChatGPT across five primary data sources on GitHub, specifically targeting \textit{Code}, \textit{Commits}, \textit{Pull Requests}, \textit{Issues}, and \textit{Discussions}, as these five sources encompass rich data and effectively capture developer interactions. Our data collection employed a hybrid approach, leveraging both GitHub's REST API \cite{GithubRest2022} and Web crawling techniques. For data accessible via direct queries, we constructed RESTful query strings tailored to the API's specifications, enabling us to search for shared ChatGPT conversation records and retrieve the relevant information. For the data in \textit{Pull Requests} and \textit{Discussions} where direct access was not feasible, we built a Web crawler~\cite{onlinepackage_TOSEM} to capture the required information.

Since OpenAI introduced the ChatGPT conversation-sharing feature in May 2023 using a standardized URL format of ``\textsc{chat.openai.com/share/\{conversation\_id\}}'', our data collection on GitHub was confined to the period from May 2023 to the collection cutoff date (June 17, 2024) when we started this work. Hence, the targeted search string is settled as ``\textsc{chat.openai.com/share}'' to retrieve the shared ChatGPT conversation records.

Due to GitHub's REST API limit of a maximum of 1,000 search results per query \cite{GithubSearchLimit}, it was not feasible to retrieve all searched entries directly from each data source. To create the most comprehensive dataset, we employed several strategies to maximize the volume of search results, adapting our approach to each of the five GitHub data sources given variations in search results (e.g., for slicing a certain time interval and programming languages). We detail our approach for collecting data from each source below:

\begin{itemize}
    \item \textbf{Code}: The majority of relevant results were from the \textit{Code} repository and most of them were invalid (e.g., incomplete URLs). Therefore, we conducted language-based searches, constructing query strings for the top 50 programming languages on GitHub \cite{GithubTopLanguages} (e.g., Python, Java, and Go), along with Markdown and HTML, which are commonly used by developers. The results from each language-specific search were then consolidated into a unified dataset.
    \item \textbf{Commits}: Regarding the ChatGPT conversation records from the \textit{Commits} repository, no additional filtering was necessary, as the total number of search results was below 1,000 at the time of data collection.
    \item \textbf{Issues and Pull Requests}: As GitHub's REST API treats both \textit{Issues} and \textit{Pull Requests} under a single ``issue'' endpoint, and the total number of search results exceeded 1,000, we employed a two-step filtering approach. First, we separated issues and pull requests using the ``\textsc{is:pr}'' and ``\textsc{is:issue}'' qualifiers, respectively. Second, we implemented a time-slicing strategy, dividing the search period from May 2023 (the release date of ChatGPT's sharing feature) to June 17, 2024 into monthly intervals. For each monthly slice, we used the ``\textsc{created}'' qualifier to retrieve issues and pull requests containing the target URL substrings. The results of all monthly slices were then combined to form a complete dataset. Then, we developed a Web crawler to extract ChatGPT conversations from the retrieved results on GitHub \textit{Pull Requests}, and the Web crawling approach was cross-verified with the results via REST API, confirming the consistency between the two approaches.
    \item \textbf{Discussions}: Due to the absence of GitHub's REST API for the \textit{Discussions} repository, we developed a Web crawler to extract ChatGPT conversation records from the retrieved results on GitHub \textit{Discussions}.
\end{itemize}


We accessed each conversation URL to collect developer-ChatGPT interactions in a prompt-response format, storing the collected data in JSON files. Table~\ref{T:Statistics} presents the statistics results: the \textbf{Initial Dataset} column reports the volume of raw data gathered from five GitHub sources, the \textit{Invalid URLs} column counts the invalid sharing of ChatGPT or GitHub links, the \textit{Non-English} column records non-English content, the \textit{Data Cleaning} column reflects the number of instances removed during data cleaning, and the \textit{Data Labeling} column presents the number of samples retained after a series of data filtering, constituting the curated dataset used for data analysis.

\begin{table}[!t]
\centering
\footnotesize
\setlength\tabcolsep{1.2pt}
\renewcommand{\arraystretch}{1.0}
\caption{Statistics of the collected data from five sources on GitHub}\label{T:Statistics}
\begin{tabular}{m{0.16\textwidth}<{\centering}m{0.14\textwidth}<{\centering}m{0.13\textwidth}<{\centering}m{0.13\textwidth}<{\centering}m{0.13\textwidth}<{\centering}m{0.13\textwidth}<{\centering}m{0.13\textwidth}<{\centering}}
\toprule
\textbf{Source} & \textbf{Initial Dataset} & \textbf{Invalid URLs} & \textbf{Non-English} & \textbf{Data Cleaning} & \textbf{Duplicates} & \textbf{Data Labeling} \\\midrule
\textbf{Code}          & 3,434 & 77   & 1,518 & 1,839 & 734 & 1,105 \\\hline
\textbf{Commits}       & 977  & 12   & 29   & 936  & 113 & 823  \\\hline
\textbf{Issues}        & 2,352 & 1,321 & 616  & 415 & 120 & 295  \\\hline
\textbf{Pull Requests} & 899  & 382  & 216  & 301 & 35 & 266  \\\hline
\textbf{Discussions}   & 176  & 77   & 0    & 99 & 41 & 58   \\\hline
\textbf{Total}         & 7,838 & 1,869 & 2,379 & 3,590 & 1,043 & 2,547 \\
\bottomrule
\end{tabular}
\end{table}

\subsection{Data Cleaning}\label{sec:Data Cleaning}
After collecting accessible data from GitHub, we performed preliminary data cleaning to eliminate invalid search results based on the following Exclusion (\textbf{E}) criteria. The results for the exclusions are presented in the \textit{Invalid URLs} and \textit{Non-English} columns of Table~\ref{T:Statistics}.

\begin{itemize}
    \item \textbf{E1. Invalid URLs}: Those invalid shared ChatGPT links could be retrieved in the initial collected data for various reasons: 
       1) \textit{Canceled Sharing}: the users who shared the conversation records may have subsequently canceled the sharing, rendering the URLs inaccessible; 
       2) \textit{Processed URLs}: the retrieval results may include processed shared ChatGPT links, resulting in entries like ``\textsc{chat.openai.com/share/xxx [invalid URL hashcode]}'', which do not correspond to valid conversation links; 
       3) \textit{Invalid URL formant}: some users shared ChatGPT conversation links with incorrect formats, such as ``\textsc{chat.openai.com/share/e/5ec5788d-8151-4c7e-ab1e-68d84a80b79f}'', in which the inclusion of ``\textsc{e/}'' deviates from the standard format. Even after removing ``\textsc{e/}'', the links remained inaccessible. Due to link disruption, we treated those instances as invalid links, as their original content could not be retrieved.
    \item \textbf{E2. Non-English text}: The conversations in the retrieved shared ChatGPT links contain non-English text (except emoji), shown in the \textit{Non-English} column of Table~\ref{T:Statistics}. Developers might use ChatGPT for tasks such as document internationalization; therefore, we excluded these non-English search results. Note that code snippets were not considered in data cleaning, meaning that we did not remove any shared ChatGPT links by their contained code snippets during data cleaning.
\end{itemize}

\begin{figure}[ht]
	\centering
        \includegraphics[width=0.85\linewidth]{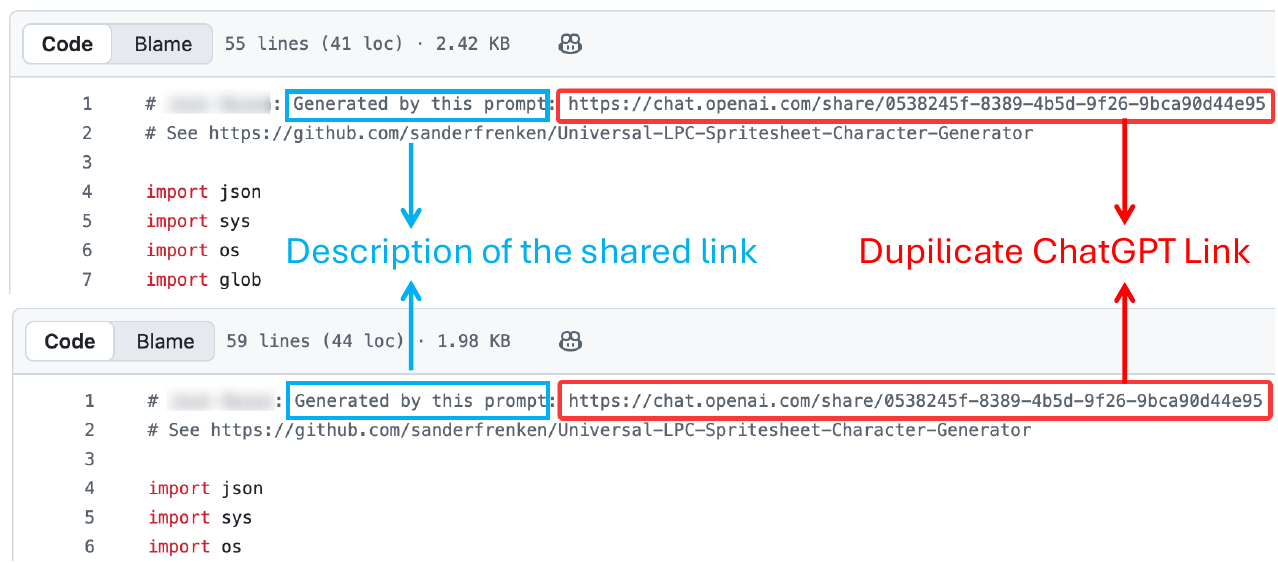}
	\caption{An example of multiple files associated with the same ChatGPT link from the \textit{Code} source}\label{F:Example}
\end{figure}

\textbf{Duplicates Removal}. After the abovementioned data cleaning by applying the Exclusion criteria \textbf{E1} and \textbf{E2}, we manually inspected the dataset and removed duplicate data. For example, developers often add or modify multiple files within a single commit; when ChatGPT links are pasted into code comments, the same ChatGPT URL can appear across several GitHub file links, as shown in Figure~\ref{F:Example}. 
Meanwhile, we have manually checked and removed the ChatGPT links that are not relevant to the content of the data sources where the links are located. Moreover, to avoid counting the same conversation multiple times, we retained only one record per unique ChatGPT link and removed the redundant records. We have conducted a deduplication procedure to eliminate records that share a single ChatGPT URL across different GitHub links, and the deduplication enhances the robustness of our results and data analysis. 

\subsection{Data Labeling}\label{sec:Data Labeling}
To streamline subsequent data labeling, extraction, and analysis, we examined developer-ChatGPT conversations in the cleaned dataset, checking the shared ChatGPT links and their contextual information within GitHub repositories. The data labeling process aims to identify and label the developer-ChatGPT interactions that are directly relevant to software development.

In terms of data labeling, all the authors discussed together to specify the data labeling task, and then the first, third, and fourth authors conducted a pilot data labeling before the formal data labeling. We used Cohen's Kappa coefficient \cite{Cohen1960} to measure the reliability and consistency of the labeled data. For the pilot data labeling, the first author reached an agreement rate of 0.79 and 0.65 with the third and fourth authors, respectively, which represents that a substantial agreement was achieved between the authors \cite{Landis1977}. All authors reached a consensus through discussions when encountering disagreements on the labels. The formal data labeling adhered to the methodology established during the pilot phase. Any discrepancies were resolved through collaborative discussions among all the authors.

\subsection{Data Extraction and Analysis}\label{sec:Data Analysis}
Our DevChat dataset was curated from five sources on GitHub, as detailed in Table~\ref{T:Statistics}. For each data source, we provided distinct metadata in JSON files to support the data analysis. Each JSON file comprises data entities with the common data items, including (1) \textit{basic GitHub repository information} (for answering RQ1) (e.g., Source \textsc{Type}, \textsc{URL}, \textsc{Author}, \textsc{Title}, \textsc{RepoName}), timestamps (e.g., \textsc{AuthorAt}, \textsc{CreatedAt}), and (2) the \textit{extracted ChatGPT data items} (for answering RQ2, RQ3, RQ4) (i.e., \textsc{ChatgptSharing}). Note that, \textsc{ChatgptSharing} consists of detailed information related to shared ChatGPT links, such as \textsc{URLs} of shared ChatGPT conversations, timestamps of the URLs (i.e., \textsc{DateOfConversation}, \textsc{DateOfAccess}), titles of ChatGPT conversations (i.e., \textsc{Title}) and corresponding content, detailed statistics of the prompts and answers (i.e., \textsc{NumberOfPrompts}, \textsc{TokenOfPrompts}, \textsc{TokenOfAnswers}), and the model version of ChatGPT used (i.e., \textsc{Model}). 

In this work, we examined how developers leverage ChatGPT throughout the software development process. For answering RQ1, we presented the distributional and usage characteristics of the shared ChatGPT links on GitHub, and we further analyzed the contents of shared ChatGPT conversations to answer the rest RQs (i.e., RQ2, RQ3, and RQ4). For the quantitative analysis addressing characteristics (RQ1) of shared ChatGPT links, we applied descriptive statistics. Concretely, we computed frequencies and percentages of the shared ChatGPT links, calculated the central tendency (mean, median, and mode) and dispersion (including the interquartile range), and visualized results distributions with histograms and bar plots. To examine and classify developers' purposes (RQ2), activities (RQ3), and tasks (RQ4) when sharing ChatGPT links, we adopted the Constant Comparison method \cite{CC2014, GT2016}, a grounded theory technique rigorously validated for deriving emergent categories from qualitative data in software engineering research. Constant Comparison was chosen specifically because it is uniquely suited for exploratory qualitative research where categories (e.g., developers' purposes) must emerge inductively from unstructured data, rather than being imposed a priori. According to the description provided in~\cite{CC2014}, the Constant Comparison process consists of three main steps: (1) \textit{initial coding}, three authors (the first, third, and fourth authors) independently reviewed the content of shared ChatGPT conversations. For example, we initially labeled 22 purposes for RQ2 before the subsequent label merge (i.e., focused coding). (2) \textit{focused coding}, the three authors selected the most frequent codes and categorized the data accordingly, with the results subsequently reviewed by the first author. For instance, we consolidated codes such as ``\textit{Dataset Split}'', ``\textit{Data Merge}'', and ``\textit{Data Cleaning}'' into ``\textit{Data Handling}'', and merged ``\textit{Program Comprehension}'' into ``\textit{Code Explanation and Understanding}''. (3) \textit{theoretical coding}, the three authors aimed to establish relationships between the identified codes. For instance, we grouped the ``Lifecycle Activities'' and ``Supporting Activities'' into the root node ``Development Related Activities'' for RQ3. To minimize personal bias, disagreements in coding results were resolved through collaborative, multi-turn discussions among all the authors, ensuring an objective interpretation of the data. Note that, we extracted and stored relevant data entities from shared ChatGPT links in MS Excel files (see the dataset at \cite{onlinepackage_TOSEM}) to facilitate data analysis. Our dataset also lends itself to further analysis of other topics, such as prompts, responses, and relevant interactions.

\section{Results}\label{sec:Results}
In this section, we present the results of our data analysis and address the RQs outlined in Section~\ref{sec:RQs}.

\subsection{RQ1: Characteristics of ChatGPT Usages}\label{sec:Characteristics}

To answer RQ1, we analyzed the characteristics of ChatGPT usage based on our curated dataset, including the distribution of shared ChatGPT links on GitHub, the associated prompts, and link descriptions.

\fakesection{(1) Data distribution of shared ChatGPT links on GitHub}

\begin{figure}[ht]
	\centering
        \includegraphics[width=0.8\linewidth]{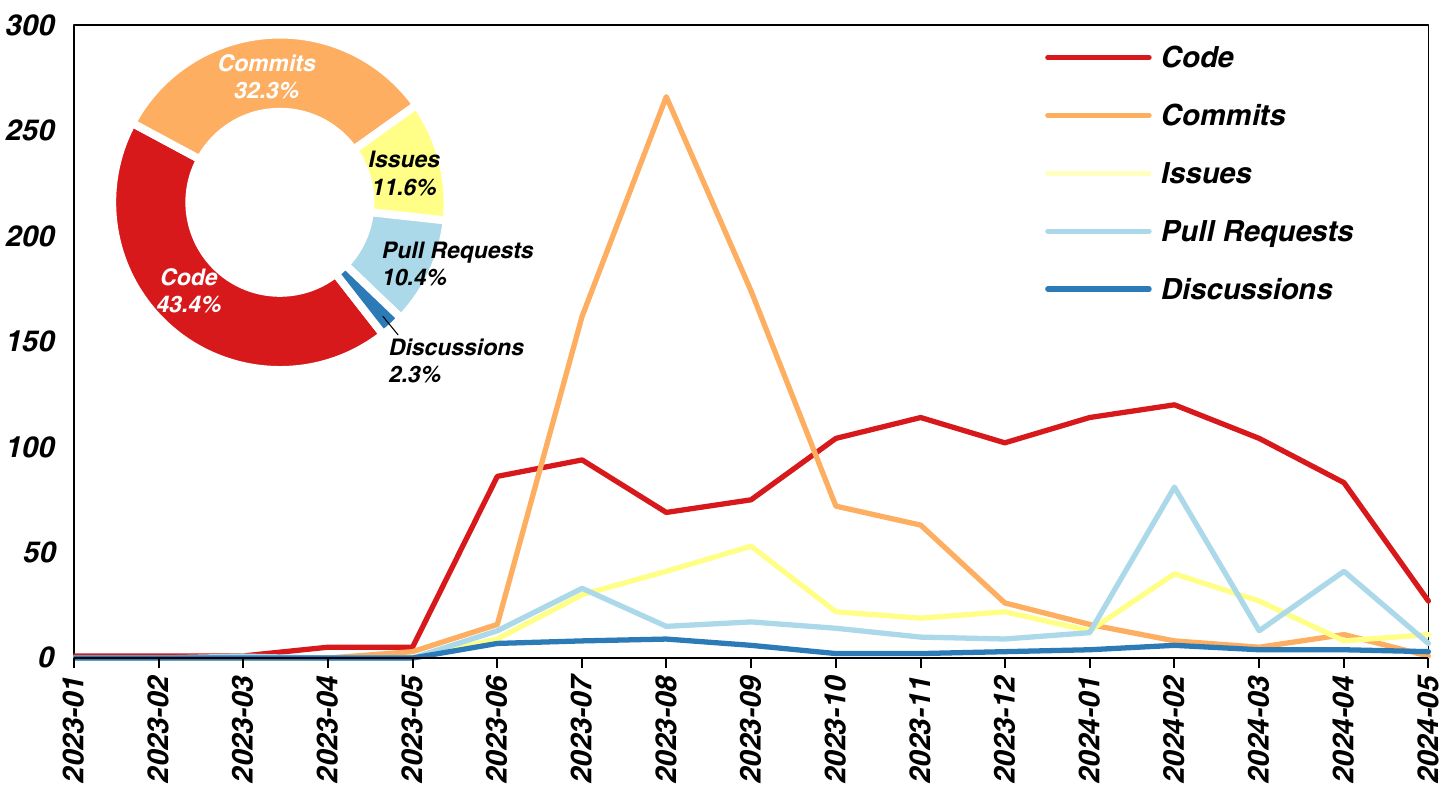}
	\caption{Data distribution of shared ChatGPT links on GitHub}\label{F:Data distribution}
\end{figure}

Figure~\ref{F:Data distribution} illustrates the distribution of 2,547 shared ChatGPT links across five data sources on GitHub (see the Donut diagram) and the monthly usage of ChatGPT by developers during software development on GitHub (see the Line diagram). This figure provides insights into the contexts where developers most commonly interact with ChatGPT within their workflow. 

Notably, \textit{Code} emerges as the dominant share of interactions, comprising 43.4\% (1,105) of all shared URLs. We can observe that developers most frequently use ChatGPT in code contexts, though the specific purposes require further analysis of the conversation content. \textit{Commits} follow as the second largest share at 32.3\% (823), reflecting that shared ChatGPT links are commonly referenced during the commit process. The remaining sources show comparatively lower usage: \textit{Issues} account for 11.6\% (295), \textit{Pull Requests} for 10.4\% (266), \textit{Discussions} for only 2.3\% (58), indicating limited reliance on ChatGPT for open-ended or collaborative discussions. Overall, the distribution shows that developer-ChatGPT conversations appear less often in \textit{Discussions} than in \textit{Code} and \textit{Commits}; this may reflect different practices regarding how ChatGPT links are shared across GitHub sources.

\begin{figure}[hbtp]
	\centering
        \includegraphics[width=0.8\linewidth]{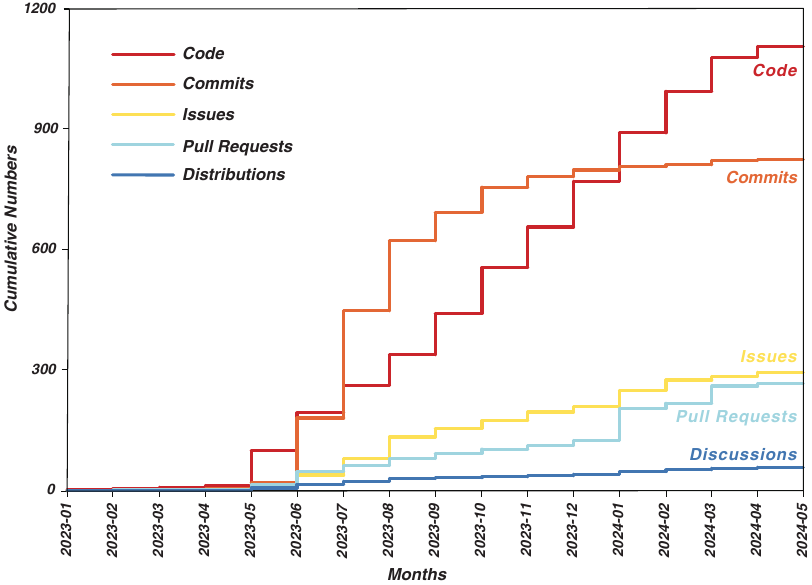}
	\caption{Cumulative usage of ChatGPT by developers during software development on GitHub}\label{F:CumulativeUsage}
\end{figure}



The Line diagram in Figure~\ref{F:Data distribution} illustrates the monthly trend of shared ChatGPT links across the five GitHub data sources from January 2023 to May 2024 (note that a few collected ChatGPT URLs were created before the release of ChatGPT's sharing feature, i.e., May 2023). Following OpenAI's release of the shared feature, a notable surge in usage occurred, peaking around August 2023, demonstrating significant initial adoption. While usage fluctuates across sources, \textit{Code} and \textit{Commits} consistently exhibit high levels of engagement, reinforcing their importance in developers' daily workflows. In contrast, \textit{Discussions} consistently exhibits lower usage, likely due to the less popularity of open-ended conversations on GitHub, and developers tend to prioritize concise, task-oriented interactions over more extensive, exploratory conversations.

From another perspective, Figure~\ref{F:CumulativeUsage} illustrates the cumulative usage of ChatGPT by developers. Specifically, the vertical axis in Figure~\ref{F:CumulativeUsage} shows the cumulative usage of ChatGPT from each source every month (see Figure~\ref{F:Data distribution}). It not only highlights the monthly growth trend but also reflects the total cumulative usage as of a specific month, which provides a more detailed view of the comparison among the five data sources. The steady growth of ChatGPT usage across the five sources highlights the platform's popularity and importance in the software development community. The lines for \textit{Code} and \textit{Commits} consistently remain the most prominent, which aligns with GitHub's primary role as a code repository and a platform for version control. The lines for \textit{Issues} and \textit{Pull Requests} show a steady increase, reflecting that collaborative development and knowledge sharing are prevalent on GitHub. The \textit{Discussions} line remains relatively low compared to the others, suggesting that discussions might not be as heavily utilized on GitHub as other activities.

Overall, Figure~\ref{F:Data distribution} and Figure~\ref{F:CumulativeUsage} provide complementary perspectives, enabling a better understanding of the data distribution characteristics from different visual angles. This distribution reflects ChatGPT's emerging role in software development, especially in collaborative and code-centric work on GitHub.

\fakesection{(2) Prompt turns distribution of shared ChatGPT links on GitHub}

\begin{figure}[hbtp]
    \centering
    \includegraphics[width=0.85\linewidth]{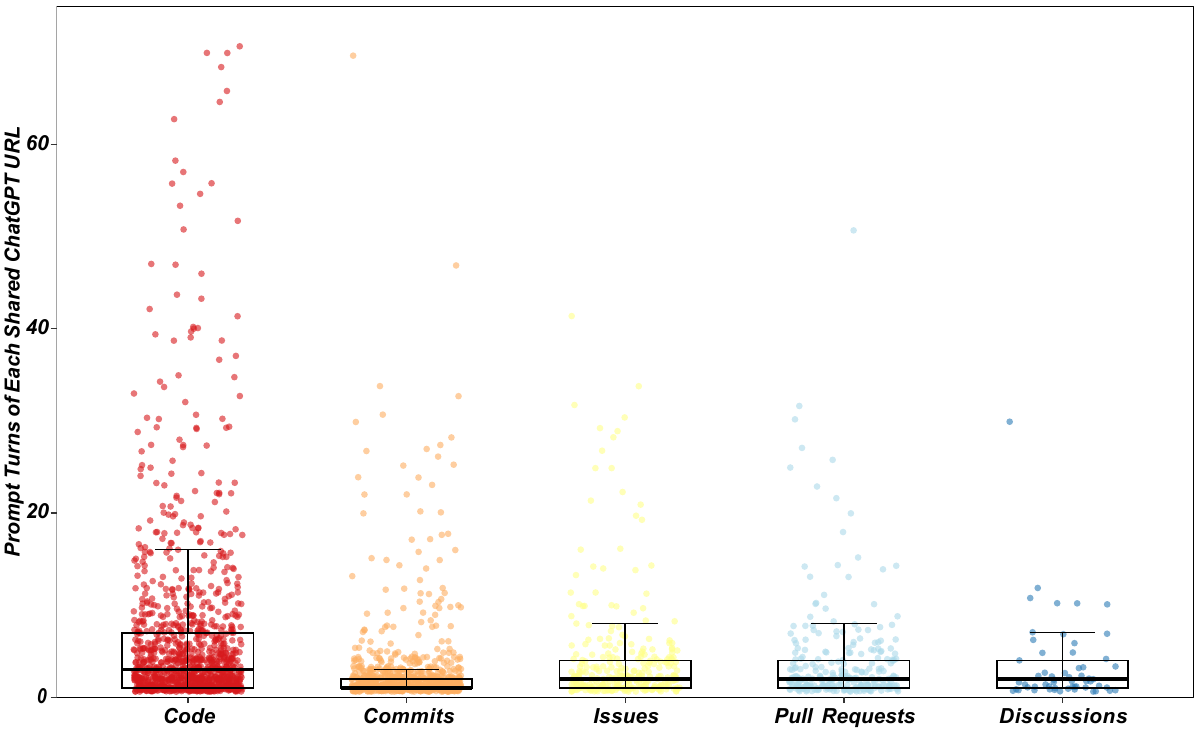}
	\caption{Prompt turns distribution of the developer-ChatGPT interactions during software development on GitHub}\label{F:Prompts}
\end{figure}

Our DevChat dataset also contains the prompt data of developer-ChatGPT interactions. In this context, a \textit{turn} refers to one complete prompt-response pair between a developer and ChatGPT (i.e., one developer's question followed by one ChatGPT's response). Figure~\ref{F:Prompts} provides a box plot visualization of the prompt turns distribution across five GitHub sources. Note that outliers (i.e., a few prompt turns over 100, see Table~\ref{T:prompts}) are not shown to ensure an appropriate scale for the coordinate axes. The corresponding statistics are summarized in Table~\ref{T:prompts}, which includes the first (Q1), second (Q2, a.k.a Median), third (Q3) quartiles of the distributions, and the InterQuartile Range (IQR), along with the mode and its corresponding frequency. Together, Figure~\ref{F:Prompts} enables a rapid visual assessment of the prompt turns distribution, and Table~\ref{T:prompts} provides a precise quantitative comparison of the prompt turn numbers.

The median number (Q2) of prompt turns in \textit{Code} interactions (3) exceeds those in other GitHub sources, with a significantly wider interquartile range (IQR = 6). The median number indicates that code-related tasks (e.g., generation, debugging, and explanation) typically involve iterative, multi-turn dialogues requiring sustained collaboration between developers and ChatGPT. Conversely, the low median values and narrow IQRs observed in \textit{Commits}, \textit{Issues}, \textit{Pull Requests}, and \textit{Discussions} reflect predominantly short and task-focused developer-ChatGPT interactions (e.g., concept interpretation, code modification).

Overall, multi-turn prompts within these interactions highlight ChatGPT's versatility, supporting not only concise tasks but also deeper discussions and collaborative problem-solving.

\begin{table}[htbp]
\centering
\small
\setlength\tabcolsep{5pt}
\renewcommand{\arraystretch}{1.0}
\caption{Statistics of the prompt turns from five sources on GitHub}\label{T:prompts}
\begin{tabular}{cccccccc}
\toprule
\textbf{Source} & \textbf{Count} & \textbf{Q1} & \textbf{Q2} & \textbf{Q3} & \textbf{IQR} & \textbf{Min/Max} & \textbf{Mode/Frequency} \\\midrule
Code          & 1,105 & 1 & 3 & 7 & 6 & 1 / 302 & 1 / 298\\
Commits       & 823   & 1 & 1 & 2 & 1 & 1 / 89 & 1 / 441\\
Issues        & 295   & 1 & 2 & 4 & 3 & 1 / 109 & 1 / 107\\
Pull Requests & 266   & 1 & 2 & 4 & 3 & 1 / 167 & 1 / 104 \\
Discussions   & 58    & 1 & 2 & 4 & 3 & 1 / 82 & 1 /25\\
\bottomrule
\end{tabular}
\end{table}

\fakesection{(3) Description of shared ChatGPT links on GitHub}

\begin{figure}[hbtp]
	\centering
        \includegraphics[width=0.7\linewidth]{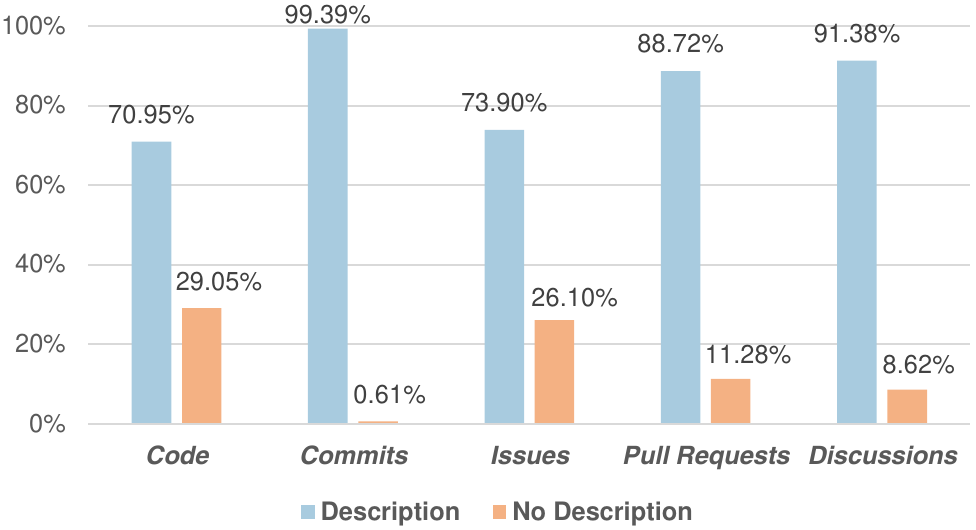}
	\caption{Description of shared ChatGPT links from the five sources}\label{F:DescriptionBar}
\end{figure}

To investigate the descriptions of shared ChatGPT links (e.g., the description in the code comment in Figure~\ref{F:Example}), we quantified the distribution of links with versus without contextual descriptions across the five GitHub sources in Figure~\ref{F:DescriptionBar}. In this study, a ``description'' refers to the accompanying textual explanation or context provided alongside a shared ChatGPT link, such as a comment, note, or sentence clarifying the purpose, content, or relevance of the linked ChatGPT interaction. We identified two primary categories when developers share ChatGPT links: sharing ChatGPT links with or without a description.

Figure~\ref{F:DescriptionBar} reveals a strong tendency to include contextual description when sharing ChatGPT links, especially in collaborations like \textit{Commits} (99.39\%), \textit{Discussions} (91.38\%), and \textit{Pull Requests} (89.72\%). In contrast, the practices of attaching descriptions to shared ChatGPT links in \textit{Code} vary, 70.95\% and 29.05\% of the shared ChatGPT links with and without a description, respectively. 
Overall, 82.80\% of shared ChatGPT links contain descriptions, whereas 17.20\% of them do not offer any contextual information. The findings reflect that developers generally prefer to provide descriptive information when sharing ChatGPT links, possibly to ensure clarity and facilitate collaborations in development with the shared developer-ChatGPT interactions.

\begin{tcolorbox}[colback=black!5, colframe=black!20, width=1.0\linewidth, arc=1mm, auto outer arc, boxrule=1.5pt]
{\textbf{RQ1 Summary}: \textit{(1) The shared ChatGPT links are predominantly from Code (43.4\%) and Commits (32.3\%), and the usage of the shared links peaked in August 2023, with consistently high engagement in Code and Commits. (2) The prompt turns distribution of shared ChatGPT links suggests that developers prefer short and task-focused interactions. (3) Most developers include contextual descriptions when sharing ChatGPT links, particularly in Commits, Discussions, and Pull Requests, but the shared ChatGPT links in Code often omit contextual descriptions.}}
\end{tcolorbox}

\subsection{RQ2: Developers' Purposes of Using ChatGPT Conversations}\label{sec:RQ2_results}

\begin{figure}[hbtp]
	\centering
        \includegraphics[width=0.9\linewidth]{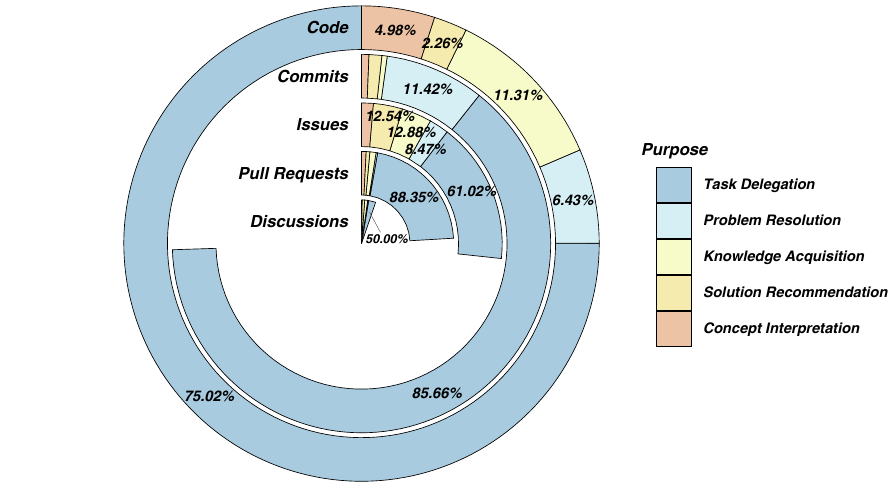} 
	\caption{Developers' purposes of using ChatGPT links from the five sources}\label{F:Purposes}
\end{figure}

To answer RQ2, we employed the Constant Comparison method \cite{CC2014, GT2016} to analyze and categorize the developers' purposes of using ChatGPT links during software development. Specifically, we examined the textual context surrounding the shared links, including the link description, following messages, and explicit user intent statements; for those links without descriptions, we analyzed their developer-ChatGPT conversations. We finally got five primary categories of developers' purposes (i.e., Task Delegation, Problem Resolution, Knowledge Acquisition, Solution Recommendation, and Concept Interpretation), and we analyzed their prevalence across the five sources. Figure~\ref{F:Purposes} presents the distribution of the five purposes across five sources. Each circle is scaled to the total number of shared ChatGPT links for different sources, and the segments within each circle show the relative contribution of each purpose, making cross-category comparisons intuitive. For each category, we provide detailed descriptions of the results, followed by one example from our curated dataset. Detailed results are presented below.

\textbf{(1) Task Delegation}: Use of ChatGPT to automate or execute specific development tasks (e.g., generating unit tests, refactoring functions). Task delegation emerges as the predominant purpose, particularly in \textit{Commits} (85.66\%) and \textit{Pull Requests} (88.35\%), reflecting ChatGPT's role in automating repetitive tasks. For instance, developers frequently employ ChatGPT to generate boilerplate code and README files. The high prevalence of task delegation in \textit{Code} (75.02\%) further underscores the utility of task delegation in accelerating code creation. However, the lower adoption of task delegation in \textit{Issues} (61.02\%) and \textit{Discussions} (50.00\%) suggests that collaborative or context-rich tasks may demand human inspection, limiting the automation potential in these sources.

\begin{tcolorbox}[colback=black!0, colframe=black!30, width=1.0\linewidth, arc=0.5mm, auto outer arc, boxrule=0.8pt, title=Example of Task Delegation, lower separated=false]
{\textbf{GitHub Link}: \url{https://github.com/hav4ik/chappity/blob/9d6d0bf90ce30089a9d421fb71fdf5fbf050d168/robot/server.py}\\
\textbf{ChatGPT Link}: \url{https://chat.openai.com/share/22827538-60fd-45d7-919c-079f22290891}\\
\textbf{Description}: This example illustrates a developer delegating a code generation task to ChatGPT, specifically to implement WebSocket classes in Python.
}\end{tcolorbox}

\textbf{(2) Problem Resolution}: Interactions aim at diagnosing and resolving faults, including debugging, fixing runtime errors, and explaining error messages. Problem resolution peaks in \textit{Commits} (11.42\%) and \textit{Issues} (8.47\%), aligning with scenarios where developers address problems when merging code or triaging issues. The minimal presence of problem resolution in \textit{Pull Requests} (1.13\%) implies that ChatGPT may not be as commonly relied upon for resolving complex or collaborative challenges. However, ChatGPT's utility in problem resolution underscores its potential as an efficient tool to identify and fix errors during development.

\begin{tcolorbox}[colback=black!0, colframe=black!30, width=1.0\linewidth, arc=0.5mm, auto outer arc, boxrule=0.8pt, title=Example of Problem Resolution, lower separated=false]
{\textbf{GitHub Link}: \url{https://github.com/CallumC1/Interact/tree/0d299d027b10d31ed76bcaa7f04b68172d6f9a39}\\
\textbf{ChatGPT Link}: \url{https://chatgpt.com/share/b5789099-c224-4a56-a3f6-169fce2052c2}\\
\textbf{Description}: This example illustrates a developer presenting an asynchronous execution problem (unexpected behavior or errors), and seeking troubleshooting advice and potential solutions from ChatGPT.
}\end{tcolorbox}

\textbf{(3) Knowledge Acquisition}: Proactive requests for clarification, background information, or domain-specific knowledge that help developers reason about design or implementation choices. Knowledge acquisition is relatively prominent in \textit{Discussions} (20.69\%) and \textit{Issues} (12.88\%), where developers seek clarifications or compare technical approaches relevant to development. While ChatGPT's role in knowledge acquisition is not as dominant in \textit{Commits} (0.61\%) and \textit{Pull Requests} (4.51\%), ChatGPT's contribution to the educational aspects of development is noteworthy. By serving as a knowledge repository and a search engine, ChatGPT helps developers quickly access relevant details and broaden their understanding of various knowledge points.

\begin{tcolorbox}[colback=black!0, colframe=black!30, width=1.0\linewidth, arc=0.5mm, auto outer arc, boxrule=0.8pt, title=Example of Knowledge Acquisition, lower separated=false]
{\textbf{GitHub Link}: \url{https://github.com/parthratra59/nomad_trakker/blob/7b2d0e364edcd98577755f9965fd8fdf6a5b69ea/src/services/operations/ProfileApi.js}\\
\textbf{ChatGPT Link}: \url{https://chatgpt.com/share/3658bbeb-0929-449f-8981-617a72165b9e}\\
\textbf{Description}: This example illustrates a developer seeking knowledge of how to set up and use Axios for API interaction, along with practical code examples for each CRUD operation.
}\end{tcolorbox}

\textbf{(4) Solution Recommendation}: Requests for options, trade-offs, design suggestions, or concrete corrective actions to guide implementation or choice among alternatives. Solution Recommendation is comparatively more frequent in \textit{Discussions} (17.24\%) and \textit{Issues} (12.54\%), indicating ChatGPT's value in collaborative problem-solving. Developers tend to use ChatGPT to generate actionable suggestions during ideation, such as recommending design patterns, best practices, and decision-making for making sound technical decisions. The lower prevalence of this task in \textit{Code} (2.26\%) and \textit{Pull Requests} (2.63\%) implies that GitHub's usage conventions that developers focus on execution (e.g., implement predefined and specific tasks like bug fixing) rather than exploration (e.g., recommend and try new possibilities like designing algorithms) in the two sources \cite{Wessel2023GApr}.

\begin{tcolorbox}[colback=black!0, colframe=black!30, width=1.0\linewidth, arc=0.5mm, auto outer arc, boxrule=0.8pt, title=Example of Solution Recommendation, lower separated=false]
{\textbf{GitHub Link}: \url{https://github.com/martyu/Rangliste/issues/4}\\
\textbf{ChatGPT Link}: \url{https://chatgpt.com/share/dde3b8ef-8c8a-4ba6-95b4-0227490d4580}\\
\textbf{Description}: This example illustrates a developer using ChatGPT for brainstorming, problem-solving, and obtaining step-by-step guidance in app development, specifically for a use case of a Schwingfest scorecard application.
}\end{tcolorbox}

\textbf{(5) Concept Interpretation}: Requests to explain, interpret, or translate specific technical concepts (e.g., algorithms, design patterns, protocols, or libraries) often at a conceptual level. Developers usually seek ChatGPT's help to understand or interpret specific concepts, algorithms, design principles, and other software development topics. Concept interpretation is the least adopted category overall, with modest usage in \textit{Issues} (5.08\%) and \textit{Discussions} (5.17\%). These requests typically highlight ChatGPT's role as a development assistant. It reflects ChatGPT's strength in facilitating concept comprehension, making ChatGPT an invaluable tool for developers seeking quick explanations or insights on unfamiliar concepts.

\begin{tcolorbox}[colback=black!0, colframe=black!30, width=1.0\linewidth, arc=0.5mm, auto outer arc, boxrule=0.8pt, title=Example of Concept Interpretation, lower separated=false]
{\textbf{GitHub Link}: \url{https://github.com/yanqi325/MovieWebsite/blob/ad1cf9b34685e93f75f18be6a7dca47a579ece44/index.html}\\
\textbf{ChatGPT Link}: \url{https://chatgpt.com/share/b80fb606-3a19-4741-844e-7f8c9265afe1}\\
\textbf{Description}: This example illustrates a developer asking ChatGPT to interpret HTML structure and styling and assist with web development topics through interactive conversations, supplemented with instances.
}\end{tcolorbox}

\begin{tcolorbox}[colback=black!5, colframe=black!20, width=1.0\linewidth, arc=1mm, auto outer arc, boxrule=1.5pt]
{\textbf{RQ2 Summary}: \textit{Based on the shared developer-ChatGPT conversations, we identified and categorized developers' purposes of using ChatGPT links during software development into five categories. The results show that Task Delegation is the dominant purpose (75-88\% Code/Commits/Pull Requests). Problem Resolution (8-11\% in Issues/Commits) and Concept Interpretation (<6\%) indicate GenAI's context limitations. Knowledge Acquisition (21\% in Discussions) and Solution Recommendation (17\% in Discussions) thrive in collaborative ideation.}}
\end{tcolorbox}

\subsection{RQ3: Development-Related Activities in Shared ChatGPT Conversations}\label{sec:RQ3_results}
\begin{figure}[htbp]
	\centering
        \includegraphics[width=0.7\linewidth]{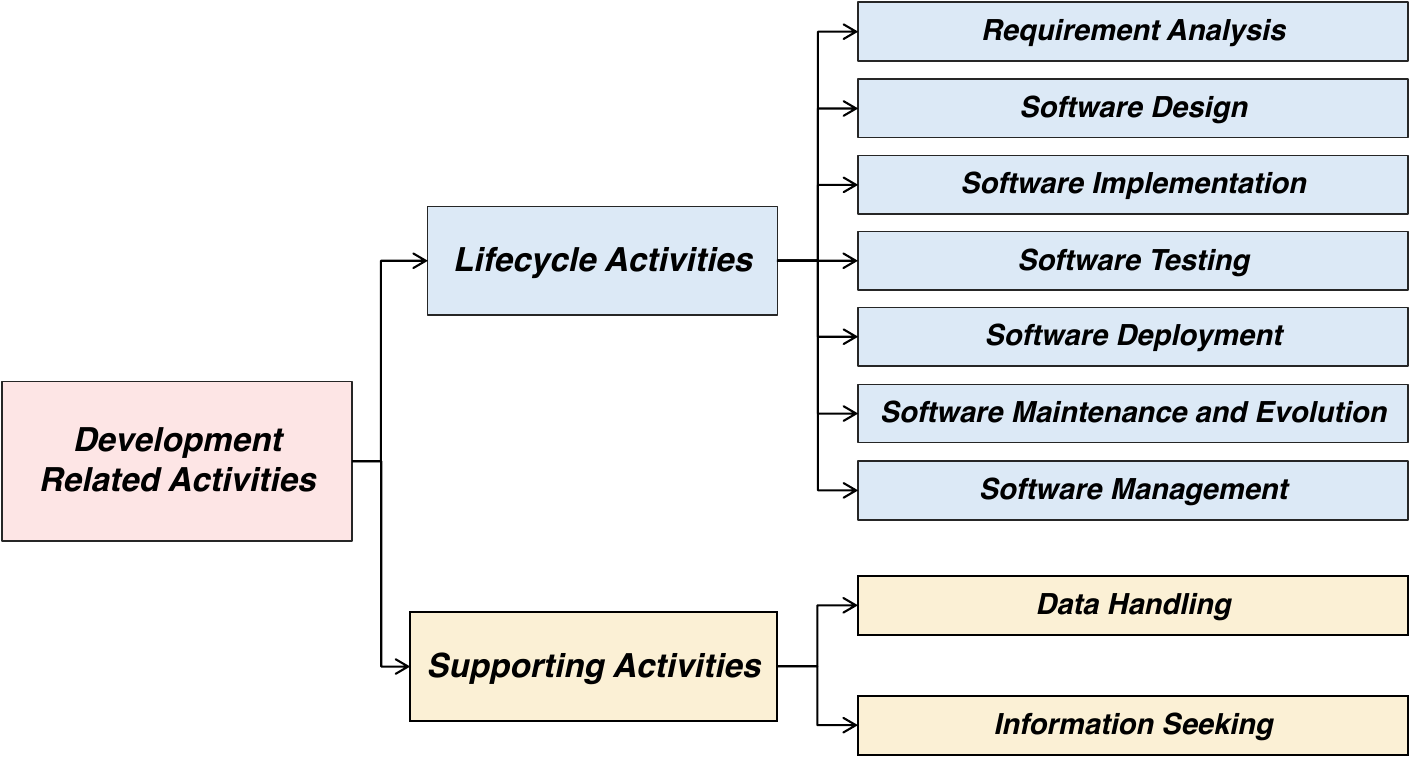}
	\caption{Categories of the development-related activities involving shared ChatGPT links}\label{F:ActivityTree}
\end{figure}

\begin{figure}[hbtp]
	\centering
        \includegraphics[width=0.97\linewidth]{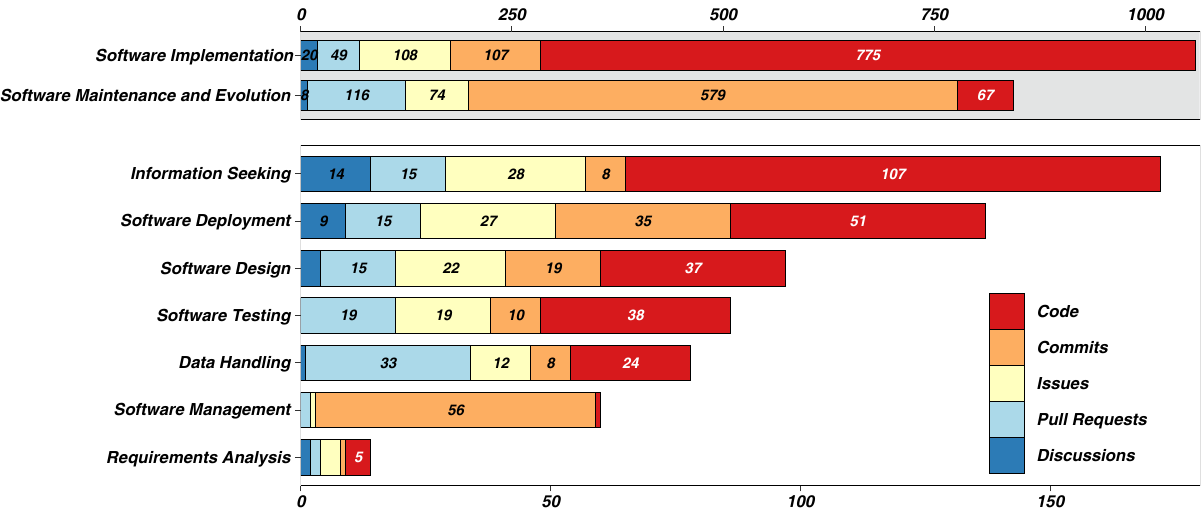}
	\caption{Distribution of development-related activities involving shared ChatGPT links from five data sources}\label{F:GroupBar}
\end{figure}

\begin{figure}[hbtp]
	\centering
        \includegraphics[width=0.9\linewidth]{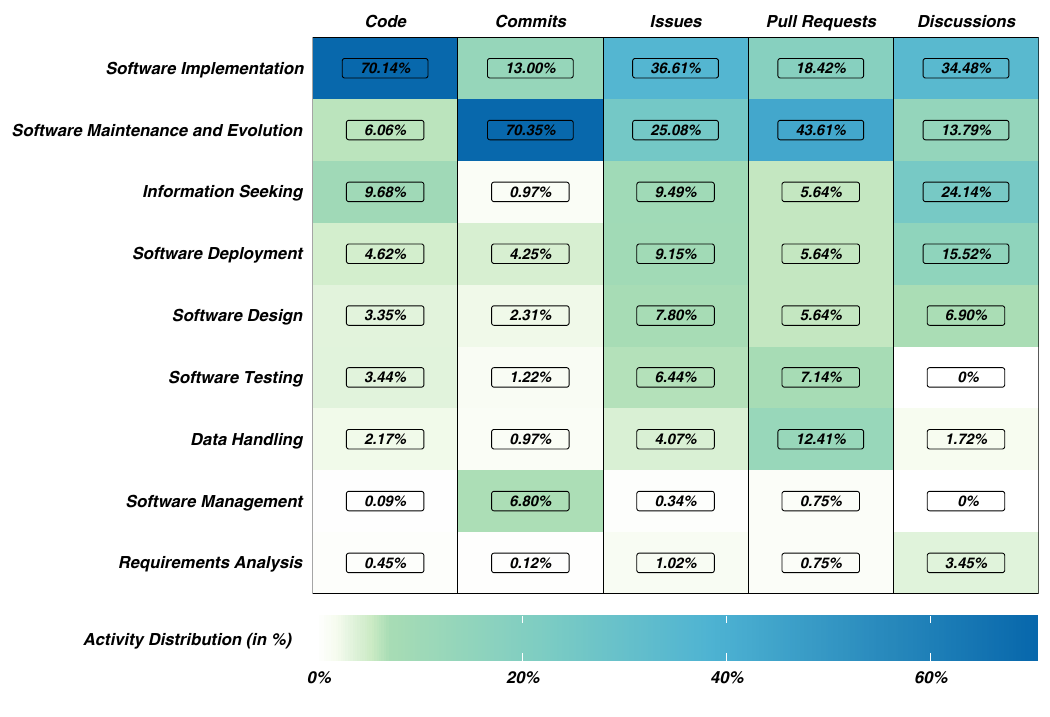} 
	\caption{Heat map of the development-related activities involving shared ChatGPT links from five data sources}\label{F:heatmap}
\end{figure}

To address RQ3, which focuses on the development-related activities (coarse-grained level) discussed in shared ChatGPT conversations, we analyzed the distribution of shared ChatGPT links across the five data sources. Specifically, we applied the Constant Comparison method \cite{CC2014, GT2016} to establish a categorization of the development-related activities, considering all extracted data items including filenames, path names, and other metadata (see Section~\ref{sec:Data Analysis}). For example, requirements and user stories are categorized as ``\textit{Requirement Analysis}''; explicit change requests and regression fixes discussions are categorized as ``\textit{Software Maintenance and Evolution}''; and path names containing ``\textsc{xxx/testing}'' or mentions of unit tests are classified as ``\textit{Software Testing}''. Besides, software documentation includes various documents which can be integrated in lifecycle activities (e.g., requirements specifications, architecture and design documents, test reports). Considering that shared ChatGPT links also serve as a form of documentation recording developer-ChatGPT interactions during software development, we chose not to treat software documentation as a separate activity. Otherwise, all the shared ChatGPT links would fall under this documentation activity. The results of RQ3 reveal the breadth of developer interactions potentially facilitated by ChatGPT, as illustrated in Figures~\ref{F:ActivityTree}, ~\ref{F:GroupBar}, and~\ref{F:heatmap}.

Although there are various ways to categorize development-related activities, we considered and adapted the classification proposed in SWEBOK \cite{SWEBOKv04} and previous studies \cite{Hou2024LLM4SE, Fan2023LLMSurvey}, which are widely recognized in the field of software engineering. Following this, we categorized the relevant activities involving shared developer-ChatGPT conversations, as shown in Figure~\ref{F:ActivityTree}. The development-related activities are defined in SWEBOK \cite{SWEBOKv04}, and we grouped them into two primary categories: \textbf{Lifecycle Activities} and \textbf{Supporting Activities}. Regarding \textit{Lifecycle Activities}, note that, activities from prior studies were not directly adopted. Instead, we identified lifecycle activities from prior studies \cite{SWEBOKv04, Hou2024LLM4SE, Fan2023LLMSurvey}, and retained only those consistently observed in our dataset \cite{onlinepackage_TOSEM}. Specifically, \textit{Requirements Analysis}, \textit{Software Design}, \textit{Software Deployment}, \textit{Software Maintenance and Evolution}, and \textit{Software Management} are the categories from \cite{Hou2024LLM4SE}, and \textit{Software Implementation} and \textit{Software Testing} are the categories from \cite{Fan2023LLMSurvey}. 

In addition, based on the labeling results, we inductively abstracted two additional \textit{Supporting Activities} (including \textit{Data Handling} and \textit{Information Seeking}), which do not map solely to a single SDLC phase but appear in ChatGPT-assisted multiple development phases. In this work, we used \emph{Development-Related Activities} as an umbrella concept to encompass all activities evidenced through ChatGPT interactions in software development contexts.

\begin{itemize}
    \item \textbf{Lifecycle Activities} encompass seven general phases of the software development lifecycle (SDLC), that is, Requirements Analysis, Software Design, Software Implementation, Software Testing, Software Deployment, Software Maintenance and Evolution, and Software Management.
    \item \textbf{Supporting Activities} include cross-phase operations (e.g., data handling) that assist and support the SDLC phases and are not tied to a specific SDLC phase. These supporting activities comprise two types of novel AI-assisted activities that facilitate and support the software development process, i.e., Data Handling and Information Seeking.
\end{itemize}

To facilitate observation and comparison, we use two scales (0\textasciitilde 1200 and 0\textasciitilde 180) in Figure~\ref{F:GroupBar}, which shows the distribution of the developer-ChatGPT conversations we collected from GitHub based on the nine activities. Notably, \textit{Software Implementation} is the most frequent activity where developers share ChatGPT conversations, accounting for a total of 1,059 (41.57\%) interactions. Most of these shared developer-ChatGPT conversations are from \textit{Code} (775, 30.43\%), \textit{Commits} (108, 4.2\%), and \textit{Issues} (107, 4.2\%), underscoring the deep involvement and widespread use of ChatGPT in modifying and improving codebases. \textit{Software Maintenance and Evolution} is the second largest development-related activity, with a total of 844 (33.14\%) interactions. Note that, in this activity, there is a high volume of \textit{Commits} (579, 22.73\%) and a substantial number of \textit{Pull Requests} (116, 4.55\%), illustrating the ongoing need for developers to refine and optimize existing software leveraging ChatGPT. Compared to the above two dominant activities, \textit{Information Seeking} is less frequent, yet still involving 172 (6.75\%) developer-ChatGPT conversations. Moreover, \textit{Software Deployment} mainly includes deployment problems in \textit{Code}, followed by moderate developer-ChatGPT conversations in \textit{Commits} and \textit{Issues}. \textit{Software Design}, \textit{Software Testing}, and \textit{Data Handling} show relatively balanced developers-ChatGPT conversations in \textit{Code}, \textit{Commits}, \textit{Issues}, and \textit{Pull Requests}. \textit{Software Management} involves ChatGPT links mainly in \textit{Commits} with \textit{Requirements Analysis} having the least developer-ChatGPT conversations. 


To provide a different and more in-depth perspective on the development-related activities involving shared ChatGPT conversations across the five sources on GitHub, Figure~\ref{F:heatmap} offers this perspective on a visual distribution heat map. This heat map depicts the percentage allocation for each data source, thereby providing detailed insights into how ChatGPT links are shared across the five sources.

\begin{itemize}
    \item \textbf{Code (1,105)}: The dominant activity where ChatGPT conversations are shared is \textit{Software Implementation}, accounting for 70.14\% of all activities, while the proportions of \textit{Software Maintenance and Evolution} (6.06\%) and \textit{Information Seeking} (9.68\%) are much lower. This implies that developers most frequently rely on ChatGPT to directly support core development tasks, such as generating and understanding code.
    Other activities, such as \textit{Software Deployment} (4.62\%), \textit{Software Design} (3.35\%), and \textit{Software Testing} (3.44\%), represent comparatively lower percentages, further reinforcing that most activities in \textit{Code} on GitHub involve direct code generation rather than less prominent activities like design or testing.
    
    \item \textbf{Commits (823)}: This source reveals a strong focus on \textit{Software Maintenance and Evolution} (70.35\%), followed by \textit{Software Implementation} (13.00\%), emphasizing ChatGPT's role in iterative improvement and code refinement through code modifications. This aligns with our intuition, given that GitHub \textit{Commits} generally indicates that developers are actively working on improving and maintaining the software, committing updates regularly to ensure software stability and improvements. Other activities like \textit{Software Management} (6.80\%) and \textit{Software Deployment} (4.25\%) contribute less, indicating relatively limited engagement with ChatGPT for software management or deployment.

    \item \textbf{Issues (295)}: \textit{Software Implementation} again ranks highest at 36.61\%, followed by \textit{Software Maintenance and Evolution} (25.08\%), indicating ChatGPT's utility in addressing core development needs like debugging and code refinement. Besides, \textit{Information Seeking} (9.49\%) is a notable activity, highlighting ChatGPT's role as a tool for knowledge retrieval within collaborative problem-solving. \textit{Software Deployment} (9.15\%) and \textit{Software Design} (7.80\%) show lower percentages, suggesting that developers also turn to ChatGPT for deployment-related troubleshooting and refining design-level decisions during issue resolution.

    \item \textbf{Pull Requests (266)}: This source is primarily related to \textit{Software Maintenance and Evolution} (43.61\%), followed by \textit{Software Implementation} (18.42\%) and \textit{Data Handling} (12.41\%). This distribution indicates that developers often turn to ChatGPT for quality assurance, feature enhancements, and data processing. Other activities show less involvement in pull requests, further highlighting ChatGPT's role as a supporting tool for ensuring software quality and consistency during code review.
    
    \item \textbf{Discussions (58)}: \textit{Software Implementation} again emerges as the leading activity (34.48\%), and followed by \textit{Information Seeking} ranks second at 24.14\%, illustrating ChatGPT's value as an on-demand knowledge source. Furthermore, \textit{Software Deployment} (15.52\%) and \textit{Software Maintenance and Evolution} (13.79\%) follow, where ChatGPT helps to troubleshoot and guide deployment. \textit{Software Design} (6.90\%) also appears occasionally, showing that discussions often involve high-level architecture and planning with ChatGPT's assistance.
\end{itemize}

\begin{tcolorbox}[colback=black!5, colframe=black!20, width=1.0\linewidth, arc=1mm, auto outer arc, boxrule=1.5pt]
\textbf{RQ3 Summary}: \textit{Analysis of shared ChatGPT links reveals that: (1) The number of shared ChatGPT links about the two lifecycle activities (i.e., Software Implementation and Software Maintenance \& Evolution) far exceeds the number of other seven activities; (2) \textit{Software Implementation}, \textit{Software Maintenance \& Evolution} are the dominant activities about which developers frequently share ChatGPT links.}
\end{tcolorbox}

\subsection{RQ4: Software Development Tasks in Shared ChatGPT Conversations}\label{sec:RQ4_results}
\begin{figure}[hbtp]
	\centering
        \includegraphics[width=\linewidth]{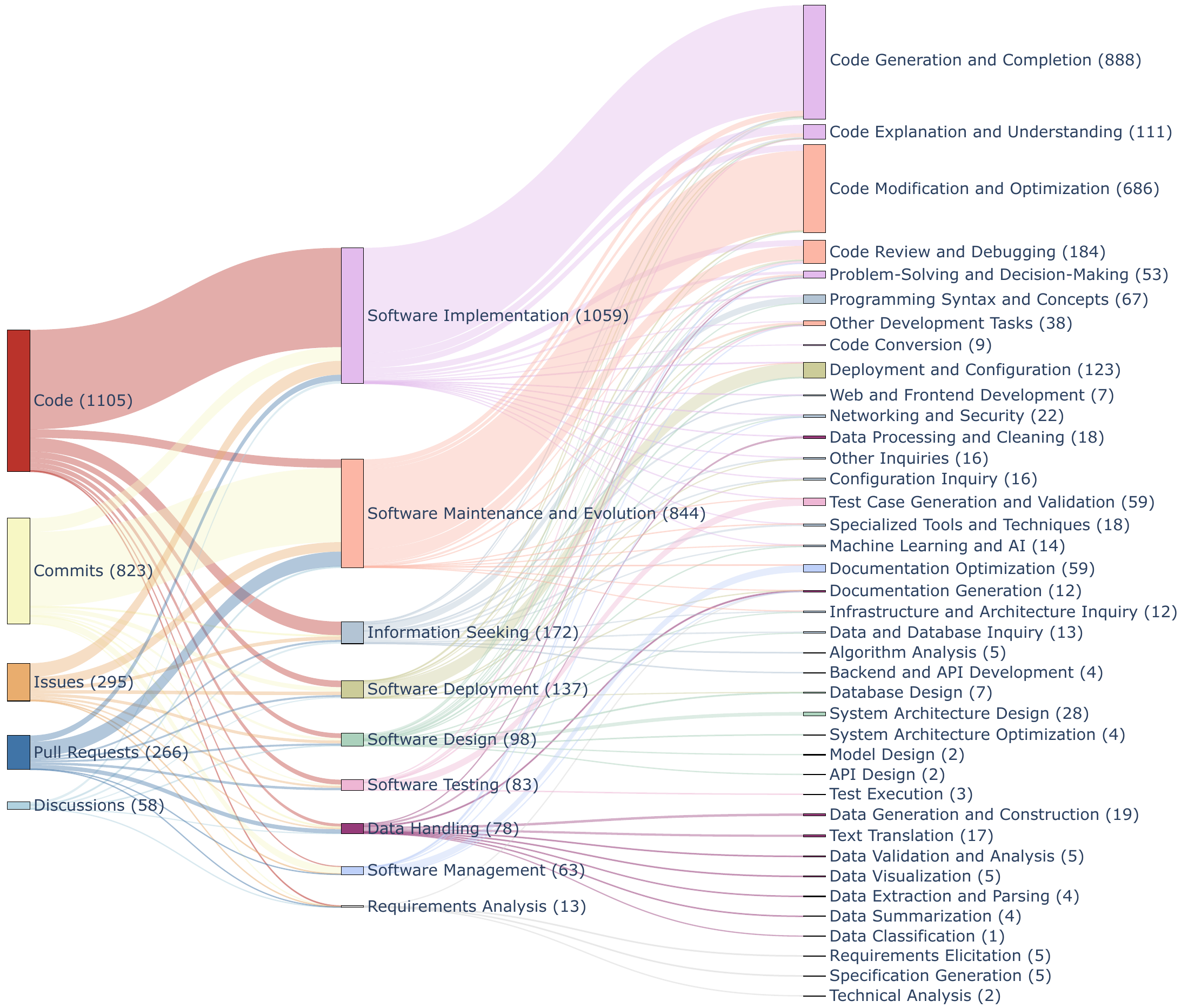}
	\caption{Mapping relationships among data sources, development-related activities, and SE tasks}\label{F:Sankey}
\end{figure}

To answer RQ4, we investigated the specific software engineering tasks (fine-grained level) within developer-ChatGPT interactions during development. Here, we adopted a broader view of \textit{tasks} by conducting an in-depth analysis of developers' intents, which encompass both action-oriented tasks (e.g., code generation) and knowledge-oriented goals (e.g., understanding specialized tools and techniques). While the latter may not constitute a task in the strict operational sense, we considered them equally relevant for characterizing developers' practical use of ChatGPT, since seeking such knowledge is often an essential step toward completing a development task. Figure~\ref{F:Sankey} presents a Sankey diagram that acts as a mapping framework linking GitHub data sources, development-related activities, and SE tasks. This decomposition visualization highlights the relationships between the five data sources, nine development-related activities, and 39 detailed SE tasks derived from ChatGPT interactions in our collected GitHub data, thereby reflecting how developers utilize ChatGPT in their daily workflows.

According to Figure~\ref{F:Sankey}, the left five data sources not only show the diversity of available information but also underscore the multifaceted nature of collaborative software development, laying the foundation for subsequent activities. The middle nodes encompass a wide spectrum of development-related activities, from requirements analysis to software maintenance and evolution. The right part of Figure~\ref{F:Sankey} breaks the nine development-related activities down into 39 specific SE tasks. In essence, the Sankey diagram visually traces the flow from each data source through the middle layer of development-related activities to specific SE tasks, revealing the deep integration of ChatGPT into the SDLC.

The most prominent flow originates from \textit{Code} (1,105) and \textit{Software Implementation} (1,059), channeling into \textit{Code Generation and Completion} (888), which reflects the frequent practice of using ChatGPT's capability to automate code generation during software development. 
Similarly, \textit{Commits} (823) strongly correlate with \textit{Software Maintenance and Evolution} (844), contributing to \textit{Code Modification and Optimization} (686) and \textit{Code Review and Debugging} (184), further reinforcing the focus on iterative code refinement. SE Tasks like \textit{Test Case Generation} (59) derive from \textit{Software Testing} (83), suggesting that ChatGPT aids in quality assurance and defect resolution.

Notably, tasks such as \textit{Deployment and Configuration} (123) and \textit{Documentation Optimization} (59) connect to \textit{Software Deployment} (137) and \textit{Software Management} (63), indicating auxiliary uses in deployment pipelines and project coordination. In contrast, underrepresented tasks like \textit{System Architecture Design} (28) and \textit{Requirements Elicitation} (5) reveal limited engagement of ChatGPT with specialized domains and early-phase development activities. This disparity also suggests developers primarily leverage ChatGPT for execution-phase tasks (e.g., code generation and completion) rather than tasks related to system design or requirements analysis (e.g., requirements elicitation).

From a macroscopic perspective, our study reveals clear usage paths of ChatGPT on GitHub. For example, most frequently, developers delegate tasks to ChatGPT during the \textit{Software Implementation} activity, accounting for the largest proportion of shared ChatGPT conversations. Within this activity, \textit{Code Generation and Completion} emerges as the most prevalent task, indicating that developers primarily rely on ChatGPT as an implementation-level assistant for producing and completing source code.

Overall, beyond illustrating the quantitative distribution of interactions, Figure~\ref{F:Sankey} offers valuable qualitative insights into the relational dynamics between data sources, development-related activities, and specific SE tasks. This visualization enables a nuanced understanding of how developers utilize ChatGPT across a broad spectrum of SE tasks, ranging from routine coding to complex problem-solving and architectural decisions. This integrated view can serve as a valuable framework for both academic research and practical applications in understanding the multifaceted nature of modern software engineering within collaborative environments.

\begin{tcolorbox}[colback=black!5, colframe=black!20, width=1.0\linewidth, arc=1mm, auto outer arc, boxrule=1.5pt]
{\textbf{RQ4 Summary}: \textit{We presented a mapping framework that links GitHub data sources, development-related activities, and SE tasks. Our analysis revealed 39 specific SE tasks across nine development-related activities, with Code Generation \& Completion and Code Modification \& Optimization emerging as the most prevalent SE tasks during software development. While code generation and maintenance tasks dominate, significant engagements were also observed in other crucial activities, highlighting ChatGPT's broad utility as an assistant throughout the SE lifecycle.}}
\end{tcolorbox}

\section{Discussions and Implications}\label{sec:Discussions & Implications}

\subsection{Interpretation of the Results}\label{sec:Discussions}

\subsubsection{Characteristics of ChatGPT Usage on GitHub}

According to the results of RQ1, the dominance of \textit{Code} (43.4\%, 1,105) and \textit{Commits} (32.3\%, 823) in shared ChatGPT interactions indicates that developers might predominantly employ ChatGPT as a technical assistant embedded in the core development tasks (e.g.,  writing, revising, debugging, and integrating code). The results also reflect that ChatGPT's role in knowledge-sharing activities such as \textit{Issue} discussions (11.6\%, 295) remains limited. The sharp increase in ChatGPT usage following the release of its sharing feature (peaking around August 2023, just three months later after the release) illustrates the rapid adaptability of developers in embracing AI tools that can be smoothly integrated into their existing workflows. In contrast, the relatively limited engagement in \textit{Discussions} suggests that developers tend to favor using ChatGPT for structured, task-specific interactions over open-ended conversations, aligning with the efficiency-focused demands of modern software development practices \cite{Vaillant2024ChatGPTsurvey}. Interestingly, the usage of ChatGPT's sharing feature peaked in August 2023, after which the growth rate began to decline. This trend likely reflects an initial surge in user engagement driven by the novelty effect following the feature's launch, which gradually tapered off as the novelty diminished.

Besides, the analysis of prompt turns distributions and descriptions regarding the shared ChatGPT links reveals a versatile integration of ChatGPT into the development process. The short and task-focused prompts (e.g., those prompts with only one or two turns) indicate that developers typically seek immediate and concise suggestions for specific issues (e.g., concept interpretation, code modification), which is crucial in fast-paced coding environments. 
Moreover, the strong tendency to include contextual descriptions of shared ChatGPT links (especially in \textit{Commits}, \textit{Pull Requests}, and \textit{Discussions}) reflects an underlying emphasis on transparency and clarity within collaborative development environments. This practice not only facilitates peer review and knowledge sharing, but also signifies a cultural shift toward embracing AI-assisted insights as an integral part of SDLC. As developers continue to employ GenAI tools like ChatGPT, these tools are well-positioned to enhance both individual efficiency and collective innovation in SE workflows~\cite{Vaillant2024ChatGPTsurvey}. 

\subsubsection{Developers' Purposes of Using ChatGPT Conversations}
ChatGPT demonstrates notable proficiency in automating routine development tasks, as reflected in the dominance of \textit{Task Delegation} within Commits (85.66\%) and Pull Requests (88.35\%). This trend points to a tangible reduction in developer workload for repetitive or procedural activities. However, ChatGPT's relatively low engagement in \textit{Problem Resolution} (11.42\%) and \textit{Concept Interpretation} (5.17\%) highlights its current limitations in handling nuanced, context-dependent challenges, thereby reaffirming the indispensable role of human expertise in complex decision-making processes (e.g., trade-off analysis considering business context and quality concerns when making architectural decisions).

Conversely, ChatGPT exhibits promising potential as a brainstorming and ideation tool, particularly in \textit{Discussions} and \textit{Issue}, in which a higher engagement is observed for \textit{Knowledge Acquisition} (20.69\%) and \textit{Solution Recommendation} (17.24\%). ChatGPT's markedly lower usage in code-centric tasks, such as Commits (0.61\%), underscores its limited reliability in scenarios demanding exactness and accountability.

The observed divergence of developers' purposes for using ChatGPT across different data sources highlights the need for future GenAI tools to be context-aware and provide tailored support aligned with specific stages of SDLC. For instance, integrating ChatGPT into issue-tracking systems could augment early-stage ideation and collaborative problem-solving, while ChatGPT's automation capabilities could be beneficial for improving pull request workflows.


\subsubsection{Development-Related Activities Involving Sharing ChatGPT Conversations}
The results of RQ3 reveal ChatGPT's dual role in supporting both SDLC activities (e.g., \textit{Software Design}) and emerging AI-augmented activities like \textit{Information Seeking} and \textit{Data Handling}. The prominence of \textit{Software Implementation} (in \textit{Code}) and \textit{Software Maintenance and Evolution} (in \textit{Commits}) highlights ChatGPT's value in accelerating the coding process and refining existing functionality. However, ChatGPT's presence across diverse data sources on GitHub, for example, aiding problem-solving in \textit{Issues}, quality assurance in \textit{Pull Requests}, and knowledge retrieval in \textit{Discussions}, demonstrates ChatGPT's adaptability to context-specific activities. This suggests that ChatGPT is not merely a code generator but a versatile assistant integrated throughout the SDLC phases to streamline SE workflows. 

Compared to the predominantly code-centric activities mentioned above, the relatively lower involvement of ChatGPT in \textit{Requirements Analysis} and \textit{Software Management} likely reflects ChatGPT's intrinsic limitations in these two activities because domain knowledge and project-specific information are still needed. As of yet, there is no doubt that such areas still rely fundamentally on human expertise and cannot be fully replaced by GenAI. Additionally, the emergence of \textit{Information Seeking} as a distinct activity indicates a paradigm shift: developers increasingly treat GenAI tools as a knowledge partner that integrates problem-solving and pair programming. This challenges traditional SDLC taxonomies, promoting a redefinition of development-related activities to account for hybrid human-AI collaboration.

\subsubsection{SE Tasks Involving ChatGPT's Engagement in Software Development}
The findings for RQ4 indicate that ChatGPT is primarily employed for the automation of coding-related tasks, specifically \textit{Code Generation \& Completion} (888 instances) and \textit{Code Modification \& Optimization} (686 instances), underscoring its value in facilitating coding and maintenance tasks. In contrast, early-phase tasks like \textit{Requirements Elicitation} (5 instances) and \textit{System Architecture Design} (28 instances) exhibit minimal usage of ChatGPT. One possible reason is that developers found ChatGPT unable to fulfill certain tasks owing to its limitations in understanding complex requirements and contributing to system design \cite{Liang2024surveyAI, Vaillant2024ChatGPTsurvey, Fan2023LLMSurvey, Hou2024LLM4SE}. Moreover, the distribution regarding the SE tasks in Figure~\ref{F:Sankey} could also confirm such a tendency: developers are predominantly using ChatGPT for on-demand assistance (e.g., coding, maintenance, troubleshooting tasks), rather than as a collaborator for architecture-level planning (e.g., architectural design and optimization). This observation raises a question about whether ChatGPT's limitations in contextual reasoning \cite{bang2023ChatGPT} and domain-specific knowledge constrain its applications in early development stages \cite{ChatGPT2025SLR}. Consequently, tasks that require intensive human judgment and holistic system understanding (e.g., \textit{Requirements Elicitation}), remain underexplored in LLM-assisted SE workflows.

In addition, ChatGPT's relatively low engagement with \textit{Software Testing} (e.g., \textit{Test Case Generation}) points to a potential gap in the adoption of LLM-powered testing automation \cite{Li2023fft}. Although LLM-generated tests could reduce manual effort, developers may distrust automatically-generated tests to accurately model complex scenarios, thereby preferring human oversight \cite{ouedraogo2024ts}. 

\subsection{Implications}\label{sec:Implications}

\noindent \faLightbulbO \quad \textbf{Establish New Paradigms for Software Development}. 

Our study results reveal how ChatGPT is being applied on GitHub, yielding usage characteristics and patterns that can help tool vendors and platform designers tailor LLM-powered features that better support developers in their daily work. Therefore, the integration of LLM-driven assistance should not merely be seen as an incremental improvement but as a fundamental shift in the SE paradigms, shaping the future of SE research, such as both development processes and decision-making in the field~\cite{Gao2024CCSE}. 

For \textit{practitioners} and \textit{tool vendors}, the usage of ChatGPT in SE tasks shows a strong focus on immediate assistance provided by ChatGPT for coding and maintenance tasks. There is significant potential for deep integration of GenAI tools like ChatGPT into developer ecosystems (e.g., IDEs, version control) to enhance context-aware support, which dynamically adapts to the developer's current code (e.g., automatically searching relevant API examples based on code context during pull-request reviews). 
Meanwhile, proposing guidelines and processes for reviewing and validating code or content generated by ChatGPT can further safeguard code quality and reliability. 

For \textit{researchers}, the findings point to promising opportunities to extend ChatGPT's utility in certain areas, such as requirements engineering, by tailoring LLMs to domain-specific SE workflows. For example, the concentration of SE tasks in \textit{Software Implementation} and \textit{Software Maintenance \& Evolution} implies that ChatGPT is primarily perceived as a coding assistant. Moreover, the findings highlight research opportunities to integrate ChatGPT in certain tasks (e.g., \textit{System Architecture Design}), thereby improving its capability to boost productivity in underexploited areas. Besides, the study underscores ChatGPT's transformative potential - not only in reshaping how developers engage in SE tasks, but also in redefining what constitutes a development-related activity in the AI era. This evolution calls for foundational knowledge in software engineering, such as SWEBOK \cite{SWEBOKv04}, to adapt alongside emerging GenAI-driven software development practices.

Nevertheless, facing the rapid advances in the field of intelligent software engineering (a.k.a AI4SE) \cite{Hou2024LLM4SE, Gao2024CCSE}, our SE research community should develop new methods, standards, and ethical guidelines that integrate intelligent AI tools or components into the SDLC, to ensure that human-AI collaboration improves code quality, developer productivity, and system maintainability~\cite{Gao2024CCSE, Terragni2024fse}.

\noindent \faLightbulbO \quad \textbf{Assess the Impact of ChatGPT on Developer Productivity and Code Quality}. 

While this study provides valuable insights regarding the daily use of ChatGPT during software development, future research could delve deeper into the impact of ChatGPT on developer productivity. In particular, controlled experimental studies could be conducted to systematically compare development outcomes, such as task completion time, code quality, and defect rates, with and without ChatGPT assistance, thereby providing more rigorous evidence of its effectiveness.

For \textit{practitioners}, understanding the impact of ChatGPT is crucial for strategic technology adoption during development. Although GenAI tools like ChatGPT exhibit great potential in enhancing developer efficiency~\cite{Vaillant2024ChatGPTsurvey}, their limitations, such as the propensity to generate inaccurate or misleading information (often referred to as \textit{hallucination} \cite{Gpt4Report2023}) or even unsafe code content~\cite{2025-CodeLM-Security}, cannot be overlooked. Therefore, for \textit{researchers}, this necessitates the collection of robust empirical evidence to assess the reliability and practical utility of ChatGPT in SE tasks. Future investigations should focus on determining the contexts in which these tools deliver substantial benefits and the contexts where they may introduce risks or inefficiencies \cite{Yu2024FFF}. Addressing these concerns will be critical to ensuring the responsible and effective integration of LLMs into modern software development.

\noindent \faLightbulbO \quad \textbf{Stay Alert on the Risks of LLM-Generated Content}. 

Our results reveal that developers utilize ChatGPT in various development-related activities (particularly \textit{Software Implementation} and \textit{Software Maintenance \& Evolution}). However, errors in ChatGPT's outputs could potentially introduce inefficiencies or vulnerabilities in software systems \cite{Liu2024cgc, Liu2024nnl}. Moreover, excessive cognitive reliance on GenAI tools might hinder the development of essential manual problem-solving skills, especially for junior developers. Given the potential of errors and biases in LLM-generated content~\cite{2025-CodeLM-Security, Xinyue2024BBLLMs}, it is imperative to establish robust safeguards, such as human oversight, rigorous validation processes, and strict adherence to ethical and legal standards. These measures are critical to maintaining the quality, security, and trustworthiness of AI-assisted development. 

Specifically, it is essential to investigate the potential risks and limitations of using ChatGPT during software development. This includes assessing concerns related to intellectual property, fairness, and the introduction of biases into LLM-generated content. \textit{Practitioners} must adopt strategies to mitigate these risks, such as incorporating human inspection for critical tasks, ensuring transparency in the LLM generation process since humans, not AI, must bear responsibility for the consequences of incorrectly generated content, and rigorously following ethical guidelines for using GenAI in software development~\cite{Akbar2023eaChatGPT}. Continuous evaluation of GenAI tools and their outputs is crucial to ensure the trustworthiness and reliability of their generated results.







\section{Threats to Validity}\label{sec:Threats}
In this section, we discuss the potential threats to the validity of our study and the adopted measures to mitigate these threats according to the guidelines of Wohlin \textit{et al}. \cite{Wohlin2012ESE}. Internal validity is not considered in this work, as no causal relationships between variables are addressed.

\textbf{Construct validity}: Potential threats to this study lie in the mined ChatGPT links and their content that developers shared on GitHub. Not all shared ChatGPT links reflect meaningful interactions; some may be noise, off-topic, or duplicates. For example, many shared ChatGPT links are not relevant to software development, or certain development-related ChatGPT links are duplicated from different sources (e.g., one link occurs in both \textit{Code} and \textit{Issues}). To mitigate these threats, we employed rigorous data cleaning techniques to eliminate noisy data (e.g., irrelevant content and duplicates) and ensure the uniqueness of the mined data through a specific ID filtering process. 



\textbf{External validity}: A relevant threat to the generalizability of our findings concerns the temporal stability of the observed findings. Our analysis focuses on developer-ChatGPT interactions on GitHub from May 2023 to June 17, 2024, a period that coincides with the release of ChatGPT's sharing feature. As GenAI models and their applications continue to evolve, the patterns of usage and perceived effectiveness may also shift over time. Additionally, the findings may not fully generalize to other open-source platforms or proprietary development environments, which may involve different user demographics or AI tools. While our study captures a significant portion of shared ChatGPT usage among developers on GitHub, it may not fully represent the broader spectrum of ChatGPT usage across all development contexts. Nevertheless, given GitHub's prevalence as the leading platform for collaborative software development, we believe that our findings reasonably capture the prevailing trends within the developer community. 

\textbf{Reliability}: The threats to the replicability of our study primarily stem from the process of data collection, cleaning, and labeling. To alleviate these threats, we have provided a detailed description of our study design in Section~\ref{sec:Study Design}, outlining the specific steps involved in data curation. Specifically, we conducted a pilot data labeling exercise to ensure a consistent understanding of the labeling criteria among the authors. Any discrepancies and disagreements were resolved through discussions. Furthermore, to promote transparency and facilitate reproducibility, we have made the datasets 
publicly available online \cite{onlinepackage_TOSEM}. This enables researchers to independently replicate our findings and conduct further research based on our shared data.



\section{Conclusions and Future Work}\label{sec:Conclusions}
The integration of LLMs, particularly ChatGPT, into software development marks a significant milestone in the evolution of SE practices. We are witnessing a paradigm shift from traditional development to AI-assisted development, streamlining SE workflows and reducing development time and effort. This study provides a comprehensive analysis of ChatGPT's practical usability and its impact on AI-assisted software development through the curated DevChat dataset, comprising 2,547 unique shared ChatGPT links from GitHub. 

Our findings reveal the characteristics of ChatGPT usage on GitHub, such as ChatGPT's prompt turns distribution, and accompanying description. Moreover, the results uncover five primary purposes for which developers share developer-ChatGPT conversations, with \textit{Task Delegation} being the dominant purpose, primarily to automate work. Furthermore, our results emphasize the widespread adoption of ChatGPT across nine development-related activities, with \textit{Code} and \textit{Commits} serving as the main data sources, and \textit{Software Implementation}, \textit{Software
Maintenance and Evolution} as the dominant activities in which developers frequently utilize ChatGPT. In addition, we established a comprehensive mapping framework to analyze the specific SE tasks for which developers employ ChatGPT during their daily routines. Overall, our study demonstrates the pivotal role of GenAI tools represented by ChatGPT in supporting a wide range of SE activities across the SDLC.

The results contribute to a comprehensive understanding of the growing impact of ChatGPT on SE practices. The results also provide actionable insights for tool builders aiming to enhance the usability of AI-assisted development tools and for researchers studying the implications of AI adoption in SE workflows. Future work could consider extending the analysis beyond the temporal scope of the DevChat dataset to examine how ChatGPT usage evolves alongside ongoing advancements in LLMs. Furthermore, researchers could study the impacts of developers' prompts on ChatGPT's outputs, and conduct comparative analyses between ChatGPT and other emerging LLMs (e.g., DeepSeek R1, Llama 3.3) to better understand their respective strengths and weaknesses. Ultimately, continued exploration of ChatGPT's role in SE will be instrumental in shaping the future of AI-driven software development practices.

\section*{Data Availability}
We have provided our collected, extracted, and processed datasets of this study online at~\cite{onlinepackage_TOSEM}.

\begin{acks}
This work has been partially supported by the National Natural Science Foundation of China (NSFC) with Grant Nos. 92582203 and 62402348, and the Major Science and Technology Project of Hubei Province under Grant No. 2024BAA008. The numerical calculations in this paper have been done on the supercomputing system in the Supercomputing Center of Wuhan University. 
\end{acks}

\bibliographystyle{ACM-Reference-Format}
\bibliography{ref}

\end{document}